\documentclass[12pt,preprint]{aastex}
\usepackage{lscape}

\slugcomment{To appear in ...}

\shorttitle{The O\,Vz Stars in GOSSS}
\shortauthors{Arias et al.}

\begin{document}


\title{Spectral classification and properties of the O\,Vz stars in the
  Galactic O-Star Spectroscopic Survey (GOSSS)}


\author{Julia I. Arias\altaffilmark{1} }
\affil{Departamento de F\'{\i}sica y Astronom\'ia, Universidad de La Serena,
  Chile}
\email{jarias@userena.cl}

\author{Nolan R. Walborn\altaffilmark{2}}
\author{Sergio Sim\'on D\'iaz\altaffilmark{3}}
\author{Rodolfo H. Barb\'a\altaffilmark{1}}
\author{Jes\'us Ma\'iz Apell\'aniz\altaffilmark{4}}
\author{Carolina Sab\'in-Sanjuli\'an\altaffilmark{1}}
\author{Roberto C. Gamen\altaffilmark{5}}
\author{Nidia I. Morrell\altaffilmark{6}}
\author{Alfredo Sota\altaffilmark{7}}
\author{Amparo Marco\altaffilmark{8,9}}
\author{Ignacio Negueruela\altaffilmark{8}}
\author{Jo\~ao R. S. Le\~ao\altaffilmark{10}}
\author{Artemio Herrero\altaffilmark{3}}
\and
\author{Emilio J. Alfaro\altaffilmark{7}}


\altaffiltext{1}{Departamento de F\'isica y Astronom\'ia, Universidad de
  La Serena, Av. Cisternas 1200 Norte, La Serena, Chile}
\altaffiltext{2}{Space Telescope Science Institute, 3700 San Martin Drive, MD
  21218, Baltimore, USA}
\altaffiltext{3}{Instituto de Astrof\'isica de Canarias, E-38200, Departamento
  de  Astrof\'isica, Universidad de La Laguna, E-38205, La Laguna,
  Tenerife, Spain}
\altaffiltext{4}{Centro de Astrobiolog\'ia, CSIC-INTA, campus ESAC, Camino
  Bajo del Castillo s/n, E-28\,692 Madrid, Spain}
\altaffiltext{5}{Instituto de Astrof\'isica de La Plata (CONICET, UNLP),
  Facultad de Ciencias Astron\'omicas y Geof\'isicas, Universidad Nacional de La
  Plata, Paseo del Bosque s/n, 1900 La Plata, Argentina}
\altaffiltext{6}{Las Campanas Observatory, Carnegie Observatories, Casilla
  601, La Serena, Chile} 
\altaffiltext{7}{Instituto de Astrof\'isica de Andaluc\'ia-CSIC, Glorieta de
  la Astronom\'ia s/n, E-18 008 Granada, Spain}
\altaffiltext{8}{Departamento de F\'{i}sica, Ingenier\'{i}a de Sistemas y
  Teor\'{i}a de la Se\~{n}al, Escuela Polit\'ecnica Superior, Universidad de
  Alicante, Carretera San Vicente del Raspeig s/n, E03690, San Vicente del
  Raspeig, Spain}
\altaffiltext{9}{Department of Astronomy, University of Florida, 211 Bryant
  Space Science Center, Gainesville, FL 32611}
\altaffiltext{10}{Universidade Federal do Rio Grande do Norte - UFRN
Caixa Postal 1524 - Campus Universitário Lagoa Nova, CEP 59078-970 | Natal/RN
- Brasil}


\begin{abstract}
On the basis of the Galactic O-Star Spectroscopic Survey (GOSSS), a detailed
systematic investigation of the O\,Vz stars is presented. The currently used
spectral  classification criteria are rediscussed, and the Vz phenomenon is
recalibrated through the addition of a quantitative criterion based on the
equivalent widths of the He\,{\sc i}~$\lambda$4471, He\,{\sc
  ii}~$\lambda$4542, and He\,{\sc ii}~$\lambda$4686 spectral lines.   
The GOSSS O\,Vz and O\,V populations resulting from the newly adopted
spectral classification criteria are comparatively analyzed. 
The locations of the O\,Vz stars are probed, showing a concentration
of the most extreme cases toward the youngest 
star forming regions. The occurrence of the Vz spectral peculiarity
in a solar-metallicity environment, as predicted by the {\sc fastwind} code,
is also investigated, 
confirming the importance of taking into account several processes for the
correct interpretation of the phenomenon. 
\end{abstract}





\section{Introduction}

Walborn (1973) first noted that the spectra of some O dwarfs in
the very young Galactic cluster Trumpler~14 in the Carina Nebula showed the
He\,{\sc ii}~$\lambda$4686 absorption substantially stronger, relative to the
other He lines, than observed in typical class V spectra.
After that, more extreme examples of objects showing this peculiarity
were found in the Large Magellanic Cloud (LMC), in particular, in the
extremely young star-forming regions N11 (Walborn \& Parker 1992, Parker et
al. 1992) 
and 30 Doradus (Walborn \& Blades 1997). These discoveries led to the
introduction, in the latter paper, of a new 
luminosity subclass named Vz, ``z'' chosen for Zero Age Main
Sequence (ZAMS), as this characteristic was hypothesized to be a
signature of youth. 

The idea behind the hypothesis that relates the 
Vz phenomenon  to the small age of the object is that 
it represents an ``inverse'' behaviour of the selective
emission effect observed among O stars, the Of phenomenon, which is due to the
emission filling of the He\,{\sc ii}~$\lambda$4686 line and correlates with
increasing 
luminosity (Walborn 2001). In other words, the typical class V spectra would
already have some emission filling in the He\,{\sc ii}~$\lambda$4686 line,
whereas the 
Vz objects would have less, thus being less luminous and less evolved,
i.e. they would be closer to the ZAMS.  

Over the last decades, numerous examples of objects belonging to the O\,Vz
class have been identified, both in the Galaxy and the Magellanic Clouds.  
Walborn (2009) presented a compilation of 50 optically observable ZAMS
O candidates that included several stars classified as O\,Vz.  
In spite of their probable key role for understanding the early
evolution of massive stars, quantitative studies of individual O\,Vz stars
have been very scarce for years. Moreover, some of these isolated studies
place the Vz 
objects on, while others locate them departed from, the ZAMS, which has
provoked some controversy regarding 
whether or not the Vz spectroscopic classification implies extreme youth.

Very recently the number of known O\,Vz stars in the LMC was significantly
increased thanks to the VLT-FLAMES Tarantula Survey (VFTS) ESO Large Programme
(Evans et al. 2011).  Its huge  
spectroscopic dataset permitted the detection of no fewer than
48 O\,Vz objects among a sample of 213 O stars in the 30~Doradus region
(Walborn et al. 2014). Sab\'{\i}n-Sanjuli\'an et al. (2014; hereafter SS14)
took advantage of this dataset to perform the first quantitative analysis of a
statistically meaningful sample of O\,Vz stars. 
Using the {\sc fastwind} stellar atmosphere code 
they obtained stellar and wind parameters for 38 O\,Vz and 46 O\,V stars, 
and found that, in general, the O\,Vz stars appear to be on or very near to the
ZAMS, although there are a non-negligible number of cases  with more advanced
ages of 2-4 Ma.  
They also investigated the predictions of the {\sc fastwind} code regarding the
Vz characteristic, and remarked on the fact that, in addition to effective
temperature and wind strength,  other stellar parameters such as gravity
and projected rotational velocity 
must be taken into account for the correct
interpretation of the phenomenon from an evolutionary point of view.
An interesting conclusion of this exhaustive research concerns the role of
metallicity in the O\,Vz phenomenon. These authors propose
that the large number of Vz stars in 30~Doradus, and the fact that some of
them are found away from the ZAMS, may be explained by the low metal content
of the LMC, which inhibits the development of a wind strong enough to break the
Vz characteristic.  As a corollary, a lower
percentage of O\,Vz stars away from the ZAMS should be expected in the Galaxy.

In the Galaxy, the Galactic O Star Spectroscopic Survey (GOSSS; Ma\'{\i}z
Apell\'aniz et al. 2011) represents the state of the art in spectral
classification of massive hot stars.  
Based on high signal-to-noise (S/N) observations from both hemispheres, it is
the largest 
collection of O-star optical spectra ever assembled. Because of the quality,
quantity, and homogeneity of the data, this survey has produced several
systemic developments and revisions, some of them unexpected, for the O-type
stars. Besides, having improved the definition of the spectral-classification
system, GOSSS has revealed numerous objects and categories of special
interest, also allowing their statistical study (Walborn et al. 2010; Sota et
al. 2011; Walborn et al. 2011; Sota et al. 2014).

The present work constitutes a further example of the potential of GOSSS.
Its unprecedented database has confirmed several previously known cases of
stars belonging to the O\,Vz class, as well as led to the discovery of many
new examples. Sota et al. (2011, 2014; hereafter S11 and S14,
respectively) present spectral  classifications for a total of 448
stars, 167 of which belong to the luminosity class V. 
Among them, 68 objects have been classified as O\,Vz. More than 150 
additional dwarfs have been observed after the publication of the first two 
papers of the project, the majority of which are included in the third 
installment (Ma\'iz Apell\'aniz et al. 2016, hereafter MA16).  Making use of
this huge amount of new data, in this paper we investigate systematically the
properties of the O\,Vz stars in the Galaxy. 
This empirical analysis is also complemented by a simple theoretical study of
the Vz phenomenon (Section~\ref{models}). 
With a limited resolving power of $R\sim2500$, the GOSSS data are not suitable
for the determination of the stellar parameters by the use of 
state-of-the-art stellar atmosphere models\footnote{In a forthcoming paper, we
  will present such a quantitative analysis using high-resolution spectra from
  two large spectroscopic surveys in the Milky Way: OWN (Barb\'a et al. 2010)
  and IACOB (Sim\'on-D\'{\i}az et al. 2011a, 2015).}.  
Then, following the procedure in the work by SS14, we used synthetic
spectra from a grid of {\sc fastwind} models 
to investigate the effect of several parameters on the occurrence of 
the Vz spectral peculiarity in a solar metallicity environment.

\section{Observations} 

All the observations used in this work come from the Galactic O-Star
Spectroscopic Survey (GOSSS). Details about the data and analysis procedures
are fully discussed in the three papers of the project (S11, S14, and MA16) 
and will not be repeated here. We recall only that GOSSS is a long-term
systematic survey of all Galactic stars ever classified as O.   
This project is providing moderate resolution ($R\sim2500$) spectroscopy in the
blue-violet region (approximately 3900-5000\,\AA), with high signal-to-noise  
ratio, typically S/N~$\sim 200-300$. The spectral types are available
through the latest version of the Galactic O-Star Catalog (GOSC, Ma\'iz
Apell\'aniz et al. 2004). In this paper we include 226 O stars from both
hemispheres pertaining to the three published GOSSS installments.

\section{Selection criteria for the sample of study}
\label{sel-crit}

\begin{figure}
\includegraphics[angle=0,scale=0.7]{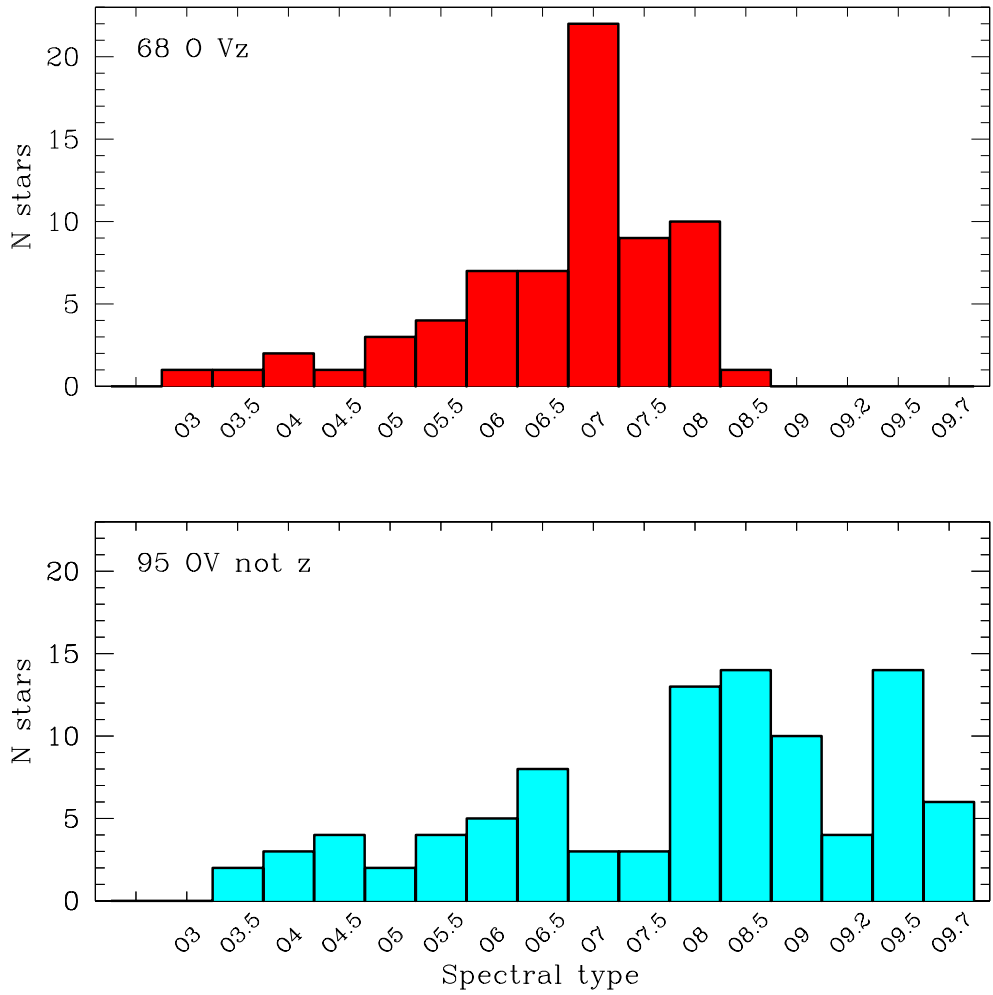}
\vspace{-100pt}
\caption{Number of stars (N stars) as a function 
  of spectral type for the Galactic O\,Vz (top) and O\,V
  non-z  stars (bottom) included in the first two papers of the GOSSS project.}
\label{dist_pub}
\end{figure}

By the beginning of this work, the GOSC contained more than three hundred 
class-V stars, spanning a spectral-type range from O3 to O9.7. Spectral types
for 163 of them had already been published by S11 and S14. We used the
original spectral classifications of the latter objects as a starting point
for the present study. 
In Figure~\ref{dist_pub} the spectral type distributions for them are
shown. The top panel shows the distribution for the 68 stars belonging, at
that time, to the Vz category, whereas the bottom panel shows that 
corresponding to the 95 objects classified as ``normal'' dwarfs, 
i.e. class V non-z. It must be stressed that these spectral classifications were
obtained 
previously to the introduction of the quantitative criterion defined in this
paper\footnote{In 
  MA16 the Vz classification scheme was adapted to the results presented in
  this paper; previous spectral classifications were thus revised and modified
  accordingly.}. 
The overall picture was the same as observed for the dwarfs in the 30~Doradus
region (see Figure~1 in SS14): 
the O\,Vz stars showed a marked concentration toward the 
intermediate spectral type O7, with a complete lack of objects at
spectral types later than O8.5, whereas 
the bulk of the normal dwarfs presented spectral types later than O8.  

%****************************************************************************
%FIGURE 1
%****************************************************************************
%****************************************************************************

The observed lack of late-type objects among the O\,Vz stars was not surprising
since,  strictly speaking, stars with spectral types later than O9 are
not relevant to the Vz phenomenon. 
As was clearly demonstrated by SS14, the Vz spectral
characteristic will never be observed at the effective temperatures typical of
these objects. Based on models computed for the metallicity of the LMC, 
these authors showed that below a certain value of $T_{\rm eff}$, 
the natural behaviour of the He lines is such that He\,{\sc i}~$\lambda$4471
is always stronger in absorption than He\,{\sc ii}~$\lambda$4686,
independently of the wind strength. 
In Section~\ref{models}, we will demonstrate that a similar argument is also
valid in the Galaxy. 
In consequence, all the stars with spectral types later than or equal to O9 
were not taken into account in the present analysis of the O\,Vz class.
It is interesting to note that, if we restrict the GOSSS published sample to
the O3-O8.5 range, 
the total numbers of objects found at that moment for the Vz and V non-z
categories were surprisingly similar (68 and 61, respectively). This means the
O\,Vz objects constituted $\sim 52\%$ of the class V stars in the relevant
spectral-type range. 

Considering the previous constraint regarding spectral types, our sample of
interest decreased 
to $223$ class V stars in the spectral-type range O3-O8.5. These objects have
been subdivided into different categories according to the following criteria.
As this study involves the measurements of the equivalent widths and
central depths of the lines present in the GOSSS spectra, we used the
appearance of the spectral lines at the relatively low resolution of GOSSS,
along with the available binarity information, in order to separate them into
different groups. 
We will go deeply into the impact of binaries in Section~\ref{binaries}, but
suffice it to say here that they may, for example, produce false  O\,Vz
spectra, and hence we had to be extremely careful when probing the binary
nature of our sample objects. 
Additionally to the multiple-epoch GOSSS spectra, information from the
literature and/or high resolution surveys was used to identify as many
binaries as possible. Then, the categories and numbers 
that characterize our sample objects are:  
(1) objects that are single lined in the GOSSS spectra, and for which no
evidence of binarity is known (132 stars listed in Table~\ref{list-singles}); 
(2) objects that are single lined in the GOSSS spectra, but are
known to be spectroscopic binaries (SBs) from high-resolution data (45
binaries, Table~\ref{list-sb1});  
(3) objects that are double lined in the GOSSS spectra (explicit SB2), 
for which the line separation is sufficiently large to allow 
measurements of central depths and equivalent widths of the individual
components by the use of deblending methods (23 binaries providing 32
components with spectral types earlier than O9); and 
(4) explicit SB2, but whose 
spectral components are not sufficiently separated to be measured
individually (15 binaries). Binaries belonging to groups (3) and (4) are
listed in Table~\ref{list-sb2}. A fifth 
category includes those objects with peculiar spectral
characteristics, for example, emission line spectra, magnetic stars,
interacting binaries (8 stars). 
In the following sections the role and significance of each of the former
categories will be explained in detail.

\section{Spectral classification criteria for the O\,Vz subclass} 

\subsection{On the problems with the central depth-based classifications}
\label{cd-ew}

%***********************************************************************
% FIGURE 2 
%***********************************************************************
\begin{figure}
\includegraphics[angle=0,scale=0.8]{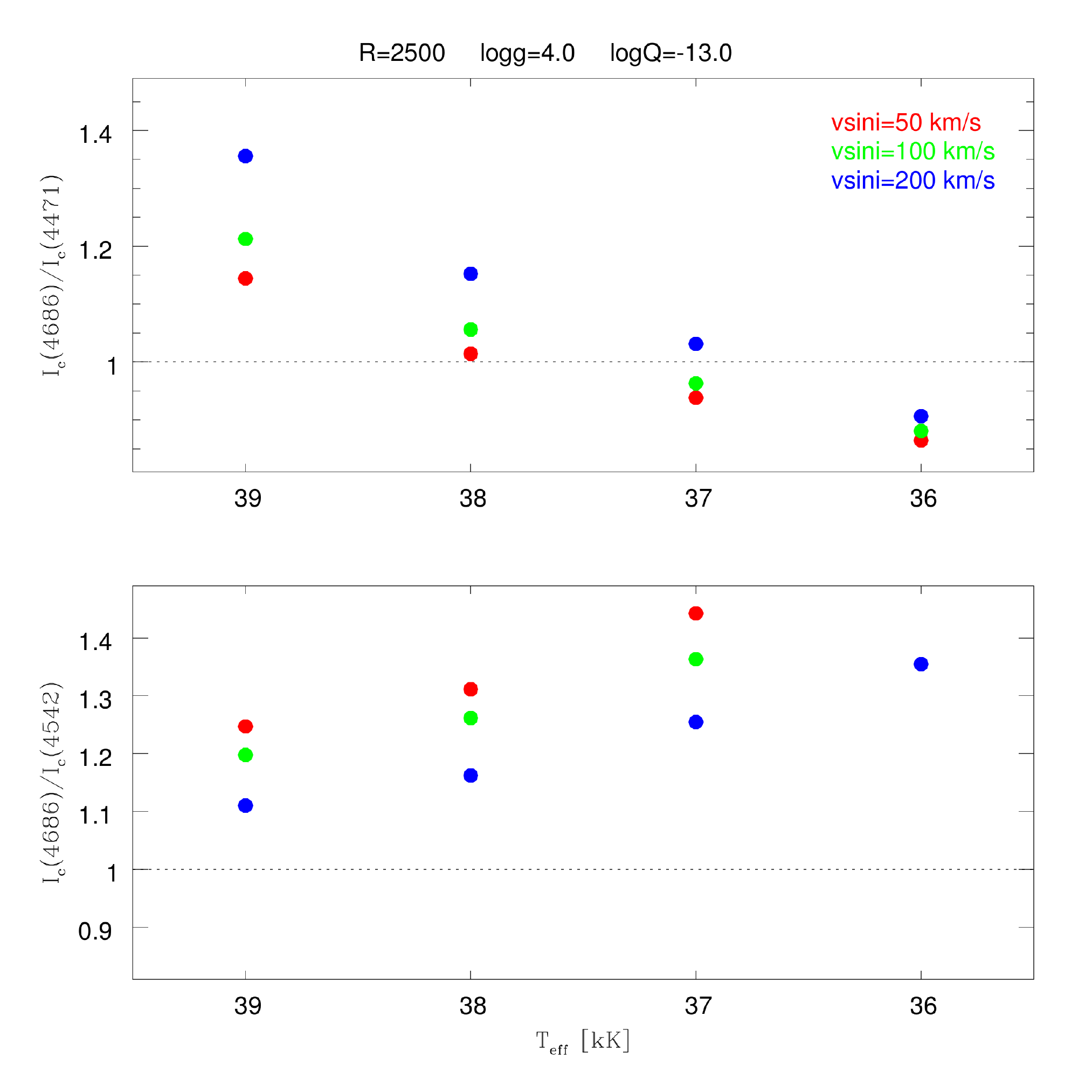}
\caption{Effect of the stellar rotation on the Vz phenomenon defined 
  from the relative depths of the He lines, as shown from {\sc fastwind} model
  predictions. The intensity $I_c$ is defined as $I_c=1-F_c$, where $F_c$ is
  the flux in the core, and indicates the central depth of the line. A star
  with effective temperature $T_{\rm eff}=37$~kK  
  (corresponding approximately to spectral type O7) may be classified either
  as Vz or   normal dwarf depending on its projected 
  rotational velocity.} 
\label{ef-rot-int}
\end{figure}
%***********************************************************************

Specifically, the current spectroscopic criterion to assign the
luminosity class Vz among O stars is that He\,{\sc ii}~$\lambda$4686
absorption is {\em stronger} than any other He line, at 
every spectral type, i. e., it is stronger than He\,{\sc ii}~$\lambda$4542 at
types earlier than or equal to O6.5, stronger than He\,{\sc i}~$\lambda$4471
at types later than or equal to O7.5, and stronger than both at type O7 where
they are equal by definition.

In the context of the morphological spectral classification scheme,  from the
advent of the digital data the former criterion is generally applied by
comparing the central depths (CDs) of the involved lines.  
However it is well known that the appearance of the spectral lines may be highly
modified by many effects, 
and that the amount and characteristics of these changes depend on
the specific line observed. 
For example, rotational broadening alters the depths of the He\,{\sc i}
and He\,{\sc ii} lines differently, 
due to the different intrinsic widths of these lines caused by the Stark
effect which is linear in He\,{\sc ii} and quadratic in He\,{\sc i}. 
In particular, the effects of rotation on the spectral types derived
morphologically are discussed by Markova 
et al. (2011). Similar effects may obviously affect the Vz
morphological spectral classification. 
Regarding this point, using {\sc fastwind} models computed for $Z=0.5\,
{\rm Z}_{\odot}$ (corresponding to the metallicity of the LMC), SS14 studied
the role of 
rotational broadening on the Vz phenomenon, finding that, {\em if we define
  the Vz characteristic from the relative depths of the He lines}, higher 
values of $v\,{\rm sin}i$ generally favour the inference of O\,Vz
classifications  at relatively low temperatures, with the opposite effect in
the high temperature regime. 

We explored the former effect using {\sc fastwind} (Santolaya-Rey et al. 1997;
Puls et al. 2005) models computed for a metallicity value of $Z={\rm Z}_{\odot}$,
reasonable for the Galactic stars studied here, and  obtained a similar result.
A convenient grid of models was chosen among the various included in the
IACOB-Grid Based Automatic Tool (IACOB-GBAT; Sim\'on-D\'iaz et al. 2011b). 
In Figure~\ref{ef-rot-int} the ratio between the CDs of
the He\,{\sc ii}~4686 and He\,{\sc i}~4471 absorptions is
plotted against effective temperature, for three values of the projected
rotational velocity, mantaining the resolution, the surface gravity, and the 
wind-strength parameter $Q$ fixed\footnote{The wind-strength parameter $Q$
  relates the mass-loss rate $\dot{M}$, the terminal velocity $v_\infty$, and
  the   stellar radius $R$, under the optical-depth invariant 
$Q = \dot{M} (R v_\infty)^{-3/2}$.}. We note, for example, that the spectrum of
a star whose effective temperature is  $37$~kK (which may reasonably
be associated to an O7 spectral type), can be 
{\em morphologically} classified either as Vz or ``normal'' class V, depending 
on the value of $v\,{\rm sin}i$.  Again, in this ``low temperature''
regime ($T_{\rm eff} \leq 37$~kK), rapid rotation increases the relevant CD
ratio, thus favouring the {\em CD-based} Vz spectral
classification.  

On the other hand, the relation between the central intensities and the
equivalent widths of the spectral lines depends on various parameters,
which leads to the following ``paradoxical'' result. Let us consider
one of the principal horizontal (i.e., spectral-class or temperature)
classification criteria for O stars, i.e. the ratio between the intensities
of the He\,{\sc i}~$\lambda$4471 and He\,{\sc ii}~$\lambda$4542 lines. By
definition, He\,{\sc i}~$\lambda$4471 is equal in strength to He\,{\sc
  ii}~$\lambda$4542 for type O7.  

\begin{figure}
\includegraphics[angle=0,scale=0.8]{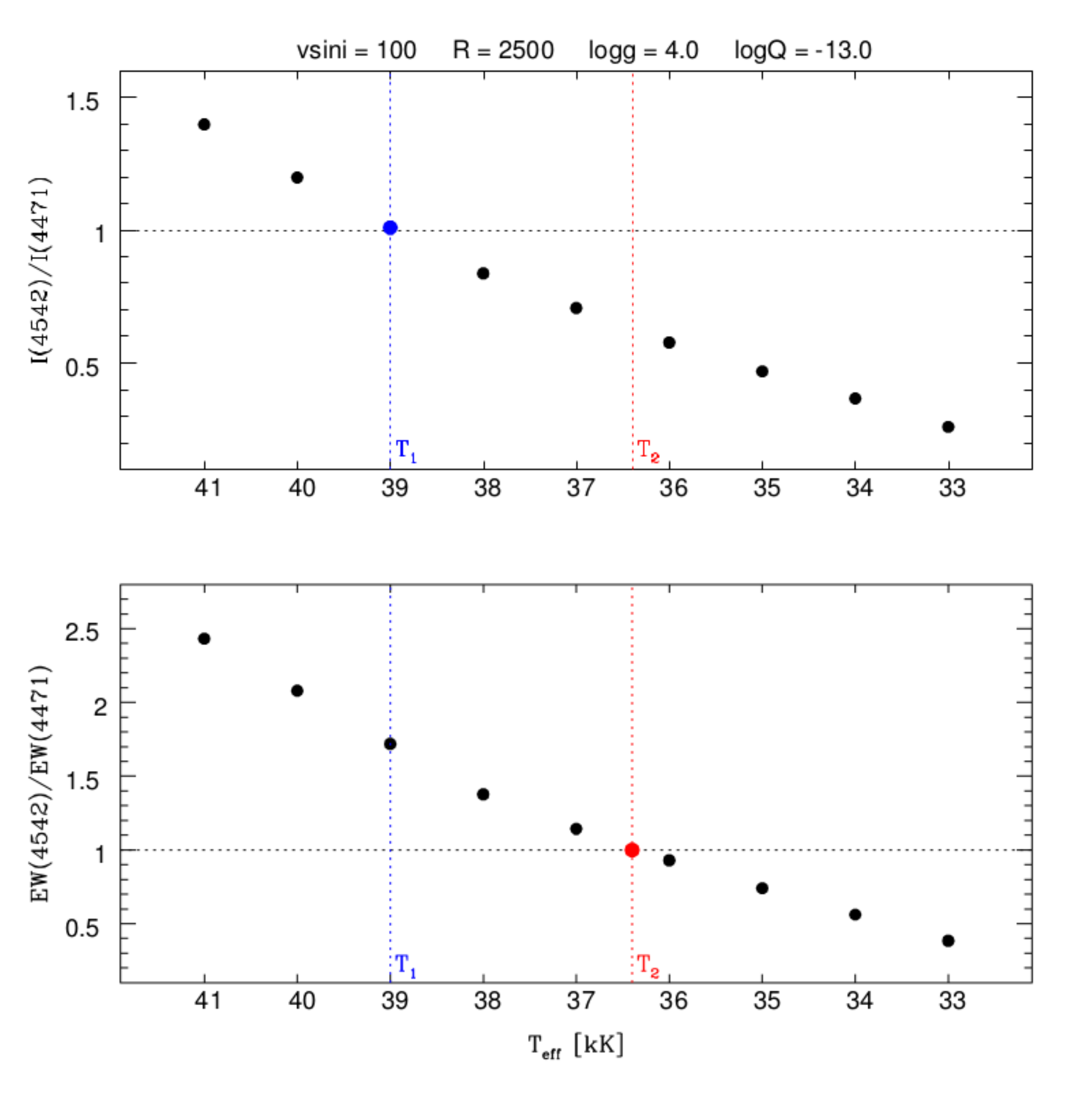}
\caption{Top panel: variation of the ratio between the CDs of the He\,{\sc
    i}~$\lambda$4471 and He\,{\sc ii}~$\lambda$4542 lines, as a function of
  $T_{\rm eff}$, as 
  shown by the {\sc fastwind} model predictions. Bottom panel: the same for the
  ratio of EWs. Fixed values of the projected rotational velocity, 
  surface gravity, resolution and wind strength have been considered. 
$T_1$ and $T_2$ indicate the effective temperature suitable for the O7 spectral type
when considered for its definition the CDs or EWs ratios, respectively.}
\label{var-CD-EW-Teff}
\end{figure}

Using {\sc fastwind} model predictions, 
we analyzed the variation of the ratios between the CDs and the EWs of the
former lines as a function of effective temperature (see
Figure~\ref{var-CD-EW-Teff}). We found that, for the considered values of
the projected rotational velocity, surface gravity, resolution and wind
strength, the CDs of the two lines are equal at $T_{\rm eff} = T_1
 \approx 39$~kK, whereas 
the  equality between their EWs occurs at the much lower temperature of
$T_{\rm eff} = T_2 \approx 36.5$~kK.  
Current calibrations (Martins, Schaerer \& Hillier 2005;  Sim\'on D\'iaz et
al. 2014), assign those temperatures to spectral types that differ by at least
one subtype.    
It then seems essential to define a consistent classification criterion.

Walborn \& Fitzpatrick (1990) stated that 
the eye is more sensitive to the EWs of the absorption lines in the
photographic spectrograms, but to the CDs in the digital 
data. Thus, systematic and random differences between the results of the two
techniques may be expected. Since EWs cannot be estimated visually in digital
data, measurements are indicated as the preferred approach.

%***********************************************************************
% FIGURE 3 
%***********************************************************************
%***********************************************************************

\subsection{A new ``quantitative'' classification criterion}
\label{crit}

Considering the above complications in using only {\em
  eye-estimated} CD ratios to define the Vz characteristic, the necessity of
taking into account an additional {\em quantitative} criterion, for example, 
based on the equivalent widths (EW) of the relevant lines, appears as very
reasonable\footnote{We must mention here that, in parallel with the
  morphological method, an alternative spectral classification 
method based on the ratios of the EW of certain lines was 
developed by Conti and collaborators, and subsequently refined
by Mathys (e.g. Conti \& Alschuler 1971, Conti \& Leep 1974, Mathys 1988, 1989). However, this quantitative system is calibrated
against the standard MK spectral types, and does not include quantitative
criteria for exceptional objects like the O\,Vz stars, which have been defined
exclusively on morphological basis.}. 
A first attempt of quantification was done by S14, where the Vz phenomenon
was considered as defined by having the ratio:

\begin{equation}
z = {\rm \frac{EW(He\,II~\lambda4686)}{Max[EW(He\,I~\lambda4471),EW(He\,II~\lambda4542)]}}
\label{def-z}
\end{equation}

\noindent greater than $\sim 1.0$. This recalibration was obtained by
measuring the EWs of the former lines in the main-sequence standard stars. 
In order to have a deeper insight into the behaviour of the z characteristic,
we have now measured the EWs and CDs of the three relevant spectral lines,
for a large sample of objects selected among the class V stars in the GOSSS
database (see Sections~\ref{sel-crit} and \ref{measurements} for details about
the sample selection). 

Because the classification categories are discrete, whereas the phenomena
are continuous, spectral classification sometimes requires interpolations or
compromises. The Vz classification is not an exception, and the situation may
be critical for these stars. Marginal O\,Vz classifications may be
generating confusion in the general picture, and might be, at least in
part, the source of the recent controversy regarding the relation
between the z characteristic and the youth of the objects. Therefore, in this
work we ``strengthen'' the definition of the Vz phenomenon by 
increasing to $1.10$ the value of the $z$ parameter defined by
equation~\ref{def-z},  just to avoid those borderline, often unclear,  cases.  
An O dwarf will then be classified as Vz when:

\begin{equation}
z = z_{4542} = {\rm \frac{EW(He\,II~4686)}{EW(He\,II~4542)}} \geq 1.10
\label{def-z-ht}
\end{equation}

\noindent at spectral types earlier or equal to O6.5, and

\begin{equation}
z = z_{4471} ={\rm \frac{EW(He\,II~4686)}{EW(He\,I~4471)}} \geq 1.10
\label{def-z-lt}
\end{equation}

\noindent at types later or equal than O7.5.
At the boundary type O7, it must occur than both $z_{4542}$ and $z_{4471}$ are
greater or equal than 1.10. 

The new critical value for the $z$ parameter was adopted by MA16 in the third
installment of the GOSSS project. Besides the reclassification of the
dwarfs from S11 and S14 (several of which have changed their   
``z status'' due to the new more strict requirement of $z \geq 1.10$), this
change implied updates on the grid of standard stars 
used for this task.
Moreover it led to the definition, for the first time, of O\,Vz standard
spectra (see Table~2 in MA16).  
Within the MK scheme of spectral classification, having a complete grid of
accurately selected standards is the key for reliable classifications.  
The quantitative condition $z \geq 1.10$ may also be used to confirm
whether or not the z qualifier must be assigned, allowing unclear marginal
cases to be objectively resolved.   

Spectral classifications and EW measurements for the presumably single
stars, single-lined binaries, and double-lined binaries included in our sample
of study are compiled in Tables~\ref{list-singles}, \ref{list-sb1}
and \ref{list-sb2}, 
respectively. 
For the three tables, column~4 contains the new GOSSS spectral classifications
taking into account the above mentioned criteria 
(reference quoted in column~5). Columns 6, 7 and 8 show the EW
measurements for the three He lines involved in the Vz phenomenon, 
and column~9 includes the $z$ parameter computed from equation~\ref{def-z}.

The spectral classifications in the three main GOSSS papers (S11, S14, and
MA16) are obtained by a combination of the classical morphological method of
visual inspection of the spectra, predominantly determined by the CDs of the
lines in the digital data,  and the use of    
MGB (Ma\'iz Apell\'aniz et al. 2012, 2015), an
interactive software designed specifically for the project. MGB compares the
observed spectrum with a grid of low-$v\sin i$ standards that can be
rotationally broadened to simulate the (n), n, nn, and nnn classification
suffixes. Therefore, MGB goes beyond 
the visual CD and line-width estimates, producing results based on the
complete profiles of the lines, and compensating the dependence of the CDs on
rotation. The results of the two techniques have been intercompared and the
final classifications adjusted slightly in some cases. In a future project
these classifications will be compared with the measured EW ratios.  We have
also verified that there is an overall good agreement between the MGB results
and the EW determinations of the z characteristic once an appropriate grid of
standards is applied.  Only for two cases, the double-lined binaries indicated
with an asterisk in Table 3, the GOSSS classifications are not in accord with
the measured $z$ parameter.

It is interesting to mention that, in the context of the present
quantification of the z  characteristic,  we became
aware of several spectra classified as type V with excessively small
values of the $z$ parameter, i.e. in the range 0.5--0.7.  A number of them %are 
were of types O4-O5.5, and inspection from this viewpoint %reveals 
revealed Of morphologies in the GOSSS data well intermediate between most
class V spectra and the class III 
standards, in terms of both the He\,{\sc ii}~$\lambda$4686 absorption and
N\,{\sc iii}~$\lambda$4640 emission strengths. This finding led to the
extension of the range of  
spectral types for which luminosity class IV is defined. The reader is
referred to paper MA16 for details about this recent change.  

\subsection{Measurements of the spectral lines}
\label{measurements}

As mentioned before, we measured the EWs 
of the three spectral lines involved in the Vz phenomenon for a large sample
of dwarfs selected from the GOSSS database.   
In Section~\ref{sel-crit} we explained that, according to the goals of the
present work, we excluded stars with spectral types later than O8.5, and
divided the $223$ remaining dwarfs in the GOSSS database into five different 
categories. With single-lined features and no evidence of binarity, 
the 132 objects included in category~1 represent the ``simplest'' case;
all of them have been measured and considered in the analysis related to the
Vz class. Their measurements are presented in Table~\ref{list-singles}.         
The 23 double-lined binaries in category~3 are also included in the
analysis. The measurements of their spectra
required the use of deblending methods. As only stars  with spectral types
earlier than O9 are relevant to the present study, only a few secondary
components, specifically 9 objects, have been measured (see for example
MY~Cam $=$ BD+56\,864 in Table~\ref{list-sb2}).   

On the other hand, being unmeasurable, none of the 15 double-lined
binaries in group~4 could be considered for the statistics. 
The peculiar objects in group~5 have not been included either in order to
avoid possible sources of noise. This category comprises a total of 8
objects.
Finally, group~2 represents a middle situation, as 33 of its 45 binaries have
been measured and included in the analysis, whereas 12 have not (see
Table~\ref{list-sb1}). 
We recall that this category comprises single-lined (at GOSSS moderate
resolution) spectra of known binary systems. 
The criterion used to discriminate the useful cases from those that should not
be considered for the statistics is the following: 
the contribution of the secondary to the composite spectrum must be
sufficiently small that the measurements obtained from the single spectral
features are representative of the primary star.
In other words, the presence of the secondary does not alter the resulting
spectral z or non-z classification.   

\begin{figure}
\includegraphics[angle=90,scale=0.6]{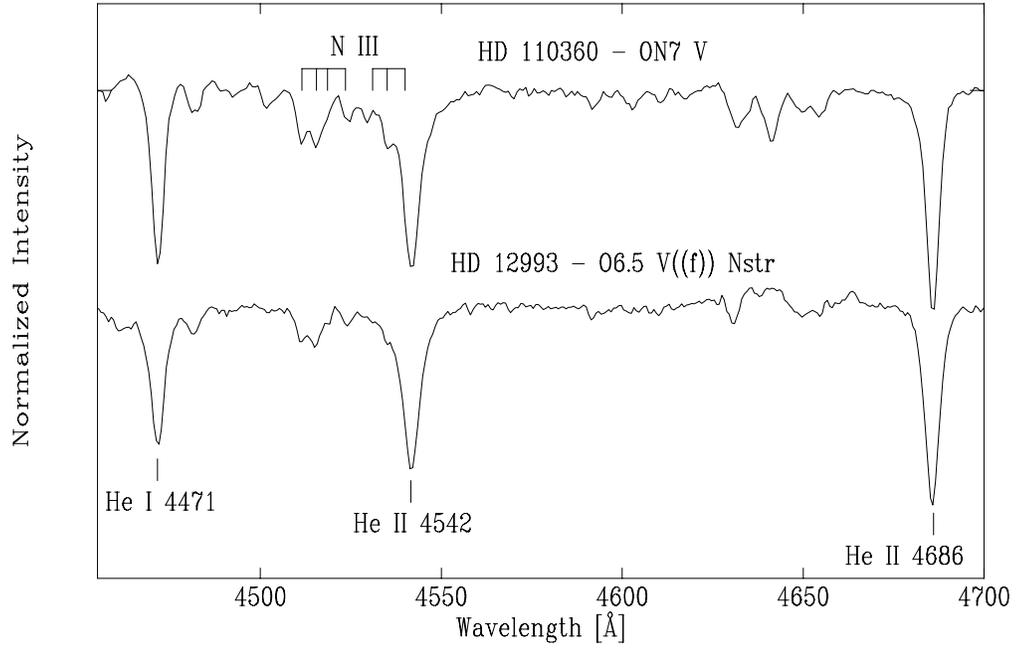}
\caption{Examples of He\,{\sc ii}~$\lambda$4542 profiles whose EW measurement requires special care, due to the blend with unusually strong N\,{\sc iii}
  lines at 4531-4535-4540\,\AA. 
  }  
\label{Nstr}
\end{figure}

To measure the EWs in our spectra, we first attempted to fit
(semi-automatically) a Voigt function to the line profiles. This method lead
to satisfactory fits for the He\,{\sc i}~$\lambda$4471 and He\,{\sc
  ii}~$\lambda$4686 lines, but 
not for the He\,{\sc ii}~$\lambda$4542 line, so the latter had to necessarily
be measured 
manually, i.e., by numerical integration of the line profile between two
points selected ``by hand'' on the continuum.  For the sake of uniformity in
methodology, we also measured manually the other two He features. 
It must be recognised that the procedure of rectifying a spectrum is 
not a trivial task, especially at low resolution, and that local rectification
errors, even small ones, 
can affect significantly the intensities of the spectral features. 
These uncertainties will impact all the methods of spectral classification,
with no exception of course for those based on EW measurements.    
In particular, the rectification of many O spectra around the He\,{\sc
  ii}~4542 line is rather complex, as this feature 
is intrinsically broader than other He lines, showing also relatively
extended wings. Additionally, in most spectra the blue wing appears 
blended with a set of N\,{\sc iii} features (4531-4535-4540\,\AA).  
Its measurement 
was thus the most sensitive. In many cases we integrated 
the red half of the profile and considered the double of that value as the best
estimate for the total EW. This effect is 
particularly important for the extreme cases of N enhanced stars, such as
HD~12\,993 and HD~110\,360, illustrated in Fig.~\ref{Nstr}. 
Note that disregarding this blend would enlarge the
EW(He\,{\sc ii}~$\lambda$4542), with a consequent decrease of the $z$ parameter
(eq.~\ref{def-z}), thus disfavouring the $z$ characteristic. 

Because of the reasons explained in the previous paragraph, the errors in the
EW measurements cannot be the same for all studied lines. For the usually
well-behaved profiles of the He\,{\sc i}~$\lambda$4471 and He\,{\sc
  ii}~$\lambda$4686 features, the mean 
absolute error of the EW can be estimated as $\sigma_{4471,4686} \leq
0.02$\,\AA.  Instead, for the more problematic He\,{\sc ii}~$\lambda$4542
line, absolute errors can be somewhat 
larger. Typical values are around 0.03\,\AA~or smaller but, to be
conservative, we adopt a mean value  $\sigma_{4542} < 0.05$\,\AA. 
The most critical cases regard the EWs of the individual components of
double-lined binaries, as deblending methods of measurement generally involve
larger errors. For those objects, absolute error values can reach
$0.1$\,\AA.    
For relative errors of 2\,\% to 7\,\% in the EW measurements, the relative error
in the $z$ parameter can be estimated in the range $5-10\,\%$. 

%***********************************************************************
% FIGURE 4
%***********************************************************************
%***********************************************************************

\subsection{The effect of binaries}
\label{binaries}

\begin{figure}
\includegraphics[angle=90,scale=0.6]{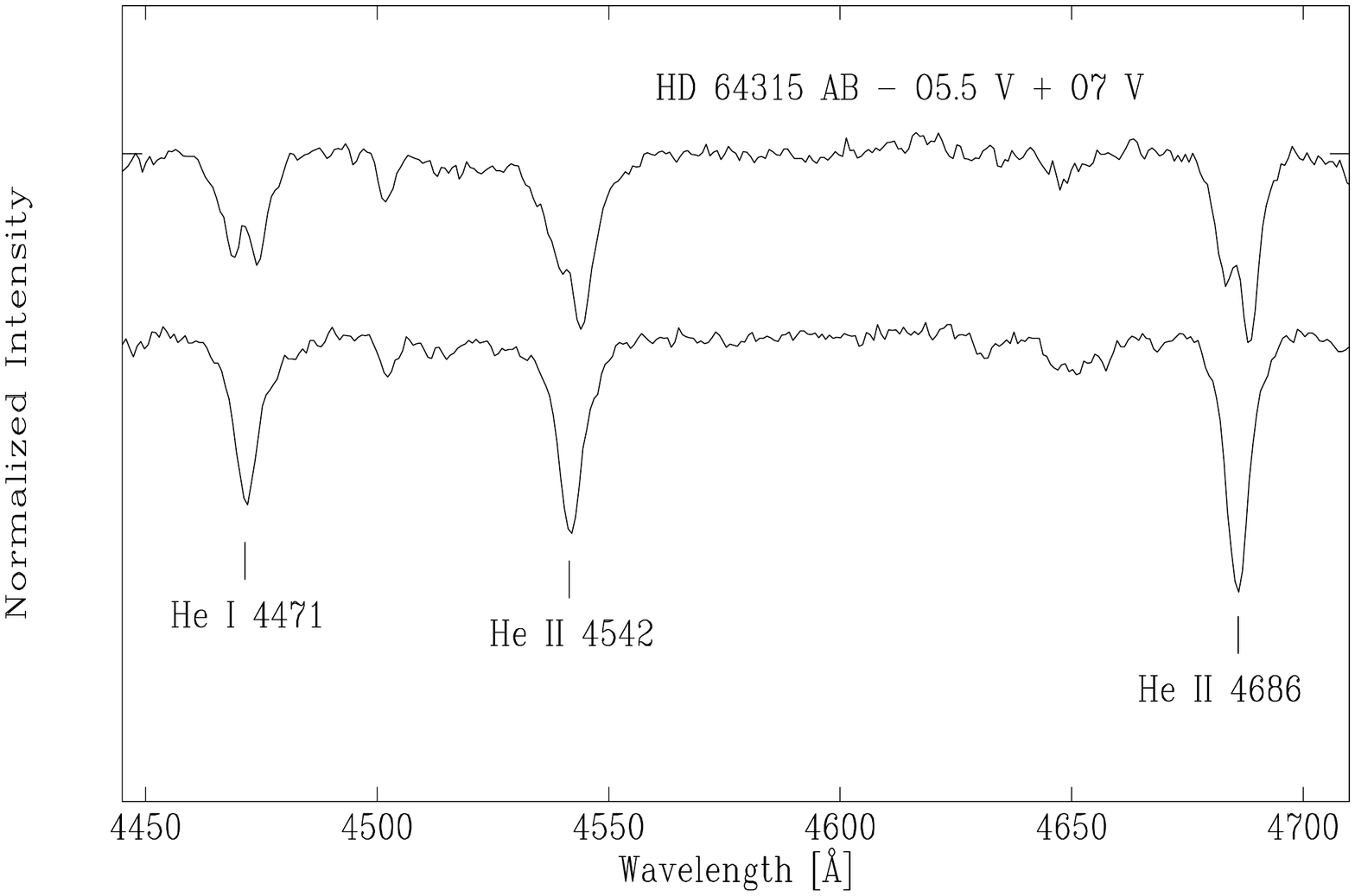}
\caption{Two spectra of the multiple system HD\,64315 illustrating how 
  binaries may produce false O\,Vz spectra. If measured during the orbital
  conjunction, the $z$ parameter of the composite spectrum is $z$=1.14. }   
\label{binary-ex-z}
\end{figure}

The high degree of multiplicity is a striking characteristic of massive stars
(Mason et al. 2009, Barb\'a et al. 2010, Sota et al. 2014, Sana et al. 2013,
2014), and also one of the most troublesome points when studying these
objects. Unknown binaries 
can easily give rise to wrong conclusions. With respect to the Vz
classification, the most critical effect 
is that they may produce false O\,Vz spectra. We have
identified a few stars from the GOSSS database 
that appear as nice intermediate O\,Vz but have subsequently been resolved
into a pair composed of an early and a late component, by means of either
other GOSSS spectra or additional high-resolution observations.  
In such an O binary, the early component dominates He\,{\sc ii}~$\lambda$4542,
and  the late component dominates He\,{\sc i}~$\lambda$4471, but both
contribute comparably to the enhanced He\,{\sc ii}~$\lambda$4686.
The case of the multiple system HD~64315 
illustrates this situation (see Figure~\ref{binary-ex-z}). This star is
composed of at least two (probably more) components (Mason et al. 2009,
Tokovinin et al. 2010, Lorenzo et al. 2010), but, if its
spectral lines  
were measured during the orbital conjunction, the obtained $z$ parameter would reach
1.14, and hence the star would be classified as Vz. 
Given the  high binary frequency among O stars, one may expect not a few
HD~64315-like cases, in which a hidden multiplicity is the actual origin 
for the observed Vz spectrum. 

%**************************************************************************
% FIGURE 5
%**************************************************************************
%**************************************************************************

We have been extremely careful when probing the binary nature of our
sample objects. Within the GOSSS database, spectra obtained in more than one
epoch are available for many stars, which allowed the identification of some
spectroscopic binaries. However, we still expected a likely important fraction
of unresolved binaries among our sample, due to the limited resolution of the
data. Therefore, in order to   identify as many binaries as possible,
we carefully searched for information from the literature, and also
took into account results from high-resolution surveys coordinated with GOSSS
like OWN and IACOB. These high-resolution monitoring programs 
appear in fact as an essential complement for a correct
interpretation of the observed statistics.

The misleading origin for some O\,Vz spectra provided by unknown binaries
may obviously be confusing the statistical interpretations. But even if they
do not produce false Vz objects, unresolved binaries may  
be ``blurring'' the statistical numbers, by leading us to count one object when
actually there are two. Among our sample of $223$ O~dwarfs, we have identified
45 objects whose spectra are single-lined at GOSSS resolution (group~2),
although they are known to be binaries from higher resolution
spectroscopic or imaging data. A recent result 
by Sana et al. (2014) claims a fraction of $100\%$ for the luminosity class V
stars that have a bound companion within 30 mas. So the key question is 
how many of the 132 stars {\em assumed} to be single in this work are
actually single. Further investigation is needed but meanwhile we must
deal with this uncertainty.

\section{The GOSSS O\,V and O\,Vz populations}
\label{sample}

Having addressed the methods applied and the problems arising in their
application, we now analyze the resulting populations of O\,V and O\,Vz stars
in the GOSC. 
    
We recall that our $223$ O dwarfs were divided into different categories
according to their binary status and 
the appearance of their spectra, and that not all of them could be measured
and included in the present analysis (see Sections~\ref{sel-crit} and
\ref{measurements}). After ``cleaning'' 
the sample, our final group of study, i.e. those stars for which a confident
value of the $z$ parameter can be derived, decreased to $132$ presumably
single stars, plus $56$ binaries or higher order 
multiple systems, $23$ of which are double-lined at the moderate GOSSS
resolution allowing individual measurements
for $9$ additional secondary components.   
A total of $197$ individual stars with spectral types earlier than O8.5 were
measured allowing us to apply both the classical 
morphological and the new quantitative criteria for their spectral
classification. We restrict our sample to this set for the statistical
considerations on 
the O\,V and O\,Vz populations. 

\begin{figure}
\includegraphics[angle=0,scale=0.8,bb=18 394 572 719]{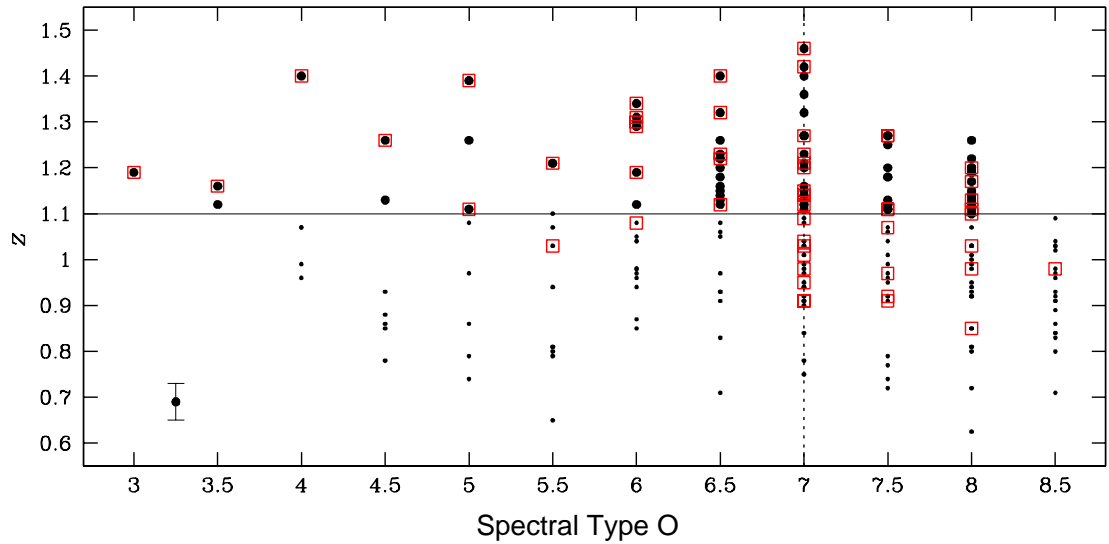}
\caption{The $z$ parameter computed from the EWs  measured in the GOSSS class-V
  spectra, as a function of spectral type. 
The red squares surround O\,Vz classified previously to the
introduction of the quantitative criterion proposed in this paper. The solid
line at $z=1.10$ indicates the current threshold to assign the z
qualifier. Only the points above this line correspond to the newly classified
O\,Vz. The representative point with error bars on the lower left corner of the
figure illustrates the typical uncertainty in the $z$ parameter.}   
\label{medidas00}
\end{figure}

\begin{figure}
\includegraphics[angle=90,scale=1.0]{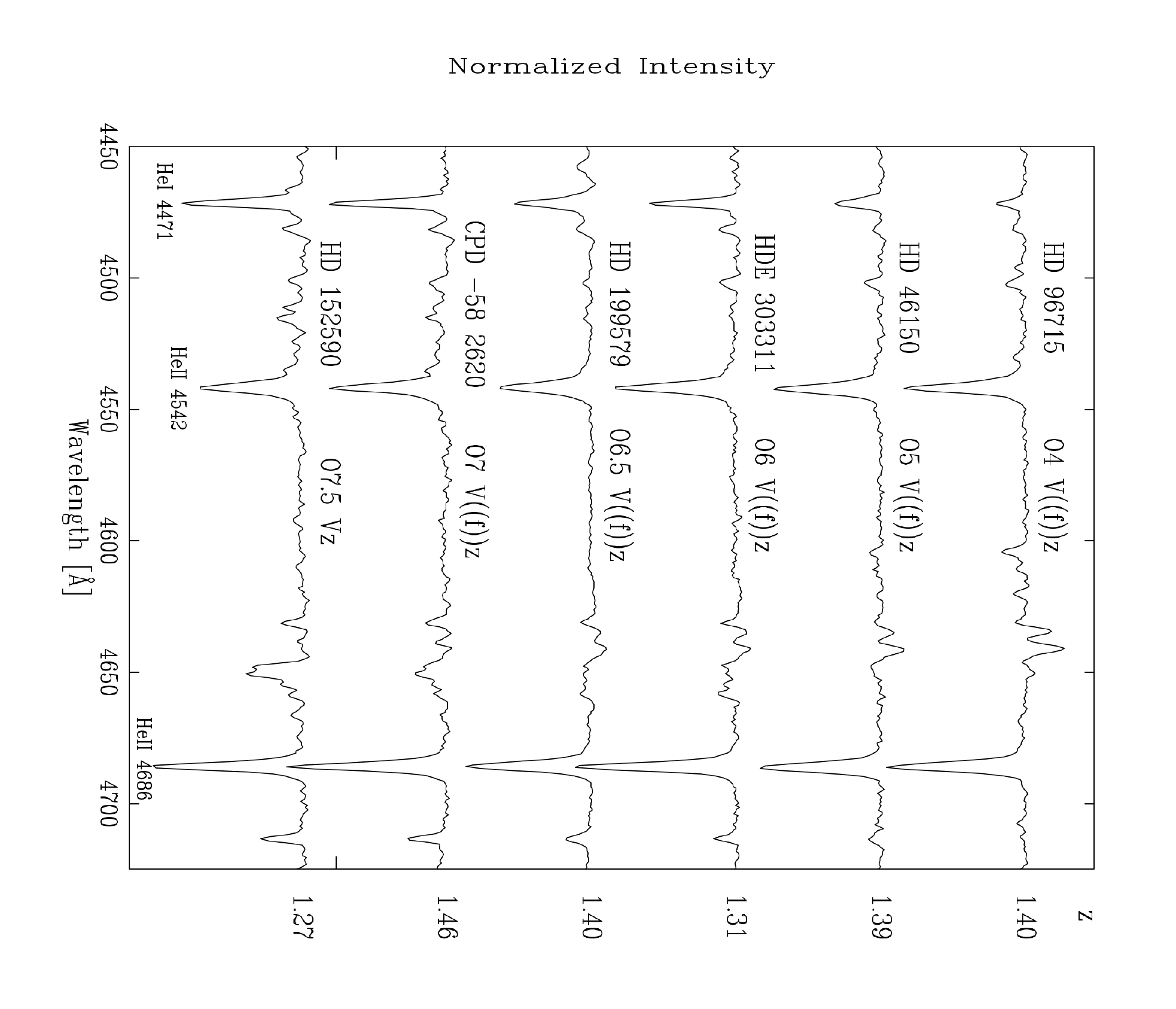}
\caption{Examples of GOSSS spectra for some of the most extreme cases
  belonging to the  O\,Vz class. The rectified spectrograms are separated by
  0.2 continuum units. The corresponding $z$ parameter is indicated to the
  right of each spectrum.}  
\label{extremas}
\end{figure}

In Figure~\ref{medidas00} the $z$ parameter computed from the EWs of the $197$
measured stars (along with additional information described afterwards) is
plotted as a function of spectral type. Note the large range of $z$ values
observed for most of the spectral types. 
The O dwarfs originally classified as Vz by S11 and S14 have been marked
with red squares. We recall the former spectral classifications were obtained
regardless of the quantitative criterion proposed in the present work. 
In the same figure,  
we have marked with a solid line the condition $z \geq 1.10$ which represents
the threshold to assign the z qualifier proposed here. Only
those O dwarfs whose $z$ parameters fall above this line are now classified as 
Vz. Slightly bigger black circles denote these cases. The representative point
on the lower left corner of the figure shows the typical uncertainty in the
$z$ parameter. Note that, for objects with $z$ close to the threshold,
measurement errors might change their ``z status''.
Some examples of spectrograms, in particular for the most extreme cases, are
shown in Fig.~\ref{extremas}.  

%************************************************************************
% FIGURE 6
%***********************************************************************
%**************************************************************************

\begin{figure}
\includegraphics[angle=0,scale=0.8]{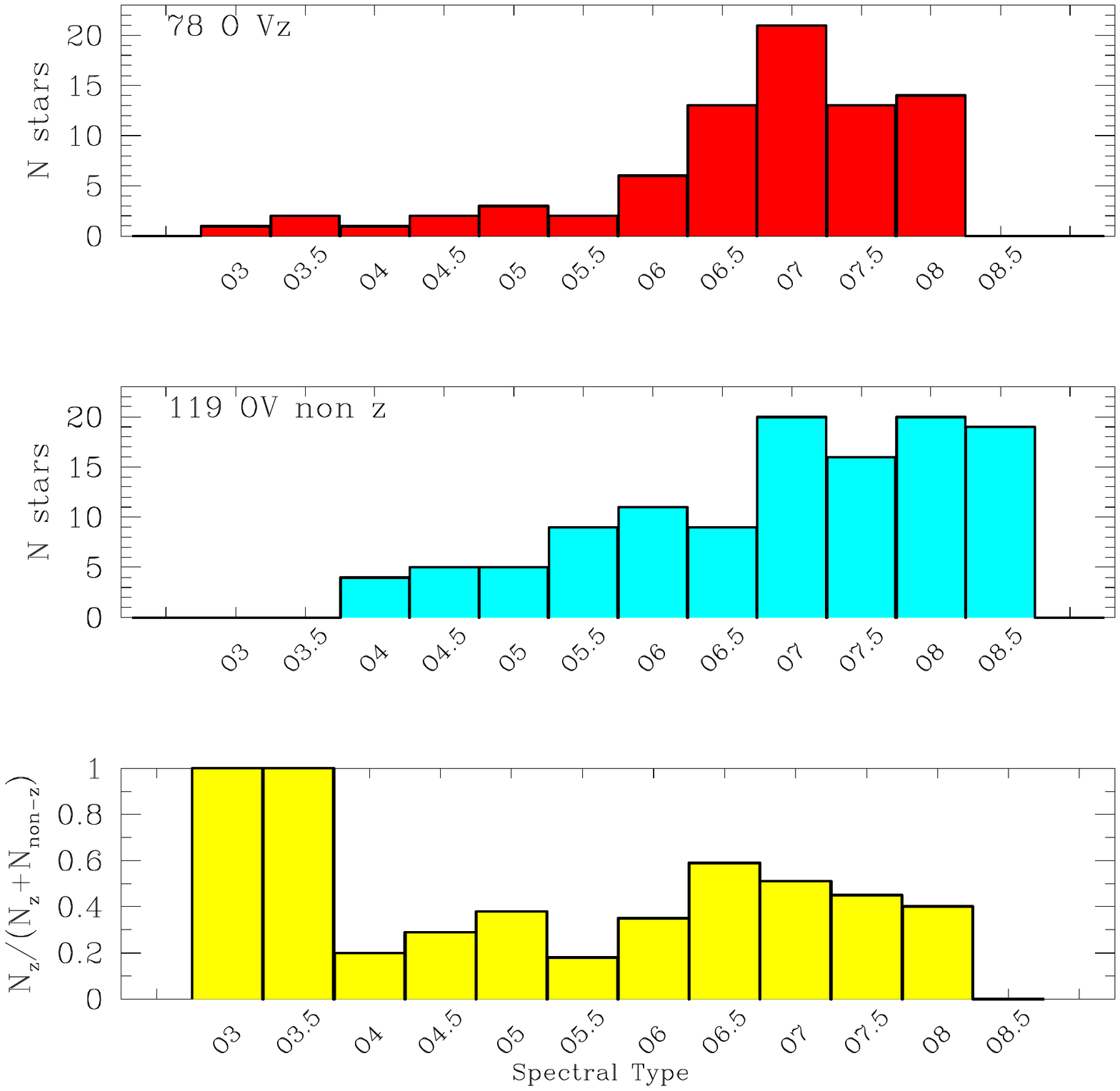}
\vspace{-150pt}
\caption{Number of stars (N stars)  as a function of spectral type for the
  Galactic O\,Vz (top panel) and 
  O\,V non-z  stars (middle panel) included in our GOSSS sample of study. The
  bottom panel shows the number of Vz objects relative to the total number of
  class-V stars at each spectral type. }
\label{sp-d}
\end{figure}

The spectral type distribution observed for the newly classified 
O\,Vz stars is presented in the top panel of Figure~\ref{sp-d}. For
comparison, an analogous distribution for the ``normal'' class V objects
spanning the same spectral type range is included in the middle panel of the
same figure.  The bottom panel shows relative differences between the two
populations for each spectral type. 
 
With the addition of the new quantitative classification criterion, the O\,Vz
population decreased to $78$ members, and thus they represent $\sim 40\%$ of the
class V stars in the relevant spectral-type range, a significantly lower
fraction, compared with the $52\%$ corresponding to  
morphology-only-based classifications. More importantly, the 
distribution for the ``normal'', i.e. non-z, dwarfs does not show that
deficiency of objects with intermediate spectral types O7-O7.5 observed in the
histograms in Figure~\ref{dist_pub}. 
Among the Vz stars however there is still a predominance of intermediate-type
objects. Specifically, 61 O\,Vz objects,  $\sim 78 \%$ 
of the whole sample, show spectral types between O6.5 and O8.   
As evident from the bottom panel of Figure~\ref{sp-d}, the populations of O\,Vz
and O\,V non-z stars are comparable in number at the former subtypes. 

We would like to compare the newly obtained spectral type distribution with
that for the O\,V stars in 30~Doradus (Fig.~1 of SS14). However the two Vz
samples were selected following different criteria, as the restrictive
quantitative condition $z \geq 1.10$ was incorporated in this paper for the
first time. A detailed comparison will make sense only after the
classification methodology is standardized. In any case, both distributions
share overall characteristics 
such as, for example, the mentioned predominance of intermediate types among
the Vz stars. 
Using {\sc fastwind} model predictions, SS14 show that part of the properties
of the O\,Vz and O\,V stars in 30~Doradus 
can be explained by a natural combination of stellar parameters. In
Sec.~\ref{models} we carry out a similar procedure to investigate the
occurrence of the Vz phenomenon (defined not from the CDs but from EWs of the
He lines) in a solar-metallicity environment. We shall show that the
distributions in Fig.~\ref{sp-d} are in good agreement with what it is
predicted by the models.

%****************************************************************************
% FIGURE 7
%****************************************************************************
%****************************************************************************

%****************************************************************************
% FIGURE 8
%****************************************************************************
%****************************************************************************

\section{{\sc fastwind} model results}
\label{models}

In a way very similar to that followed by SS14, we use synthetic spectra 
computed with the {\sc fastwind} stellar atmosphere code to investigate how
the variation of the most relevant parameters, namely, wind strength ($Q$) and
effective temperature ($T_{\rm eff}$), impact in the occurrence of the Vz
peculiarity.   
The differences with the former analysis are twofold: first, we use a 
grid of models computed for the solar metallicity $Z={\rm Z}\odot$, which is an
appropriate mean value for the GOSSS stars; and second, we consider the
quantitative criterion based on the EWs of the He\,{\sc i}~$\lambda$4471,
He\,{\sc  ii}~$\lambda$4542, and He\,{\sc ii}~$\lambda$4686 lines (see
Sec.~\ref{cd-ew}) for the 
definition of the Vz characteristic.  
The main conclusions from the present analysis are qualitatively the
same found by SS14, who were the first to point out the importance of
taking into account additional parameters to interpret correctly the Vz
phenomenon.   
However numerical results are somewhat different, most probably due to the
distinct values of the metallicity considered in each case. 

As discussed in Sec.~\ref{crit}, an O7 dwarf will be classified as Vz when
both $z_{4542}$ and $z_{4471}$ are 
greater or equal than 1.10. The effective temperature for this boundary type
is represented by $T_2$ in 
Figure~\ref{var-CD-EW-Teff}. For the values of projected rotational velocity,
surface gravity, resolution, and wind strength indicated at the top of that
figure, $T_2$ is between 36 and 37~kK. Although it depends on the
specific values considered for the former parameters, 
such a value is a reasonable 
estimate for the limit between the use of $z_{4542}$ and $z_{4471}$.    

\begin{figure}
\includegraphics[angle=0,scale=0.8]{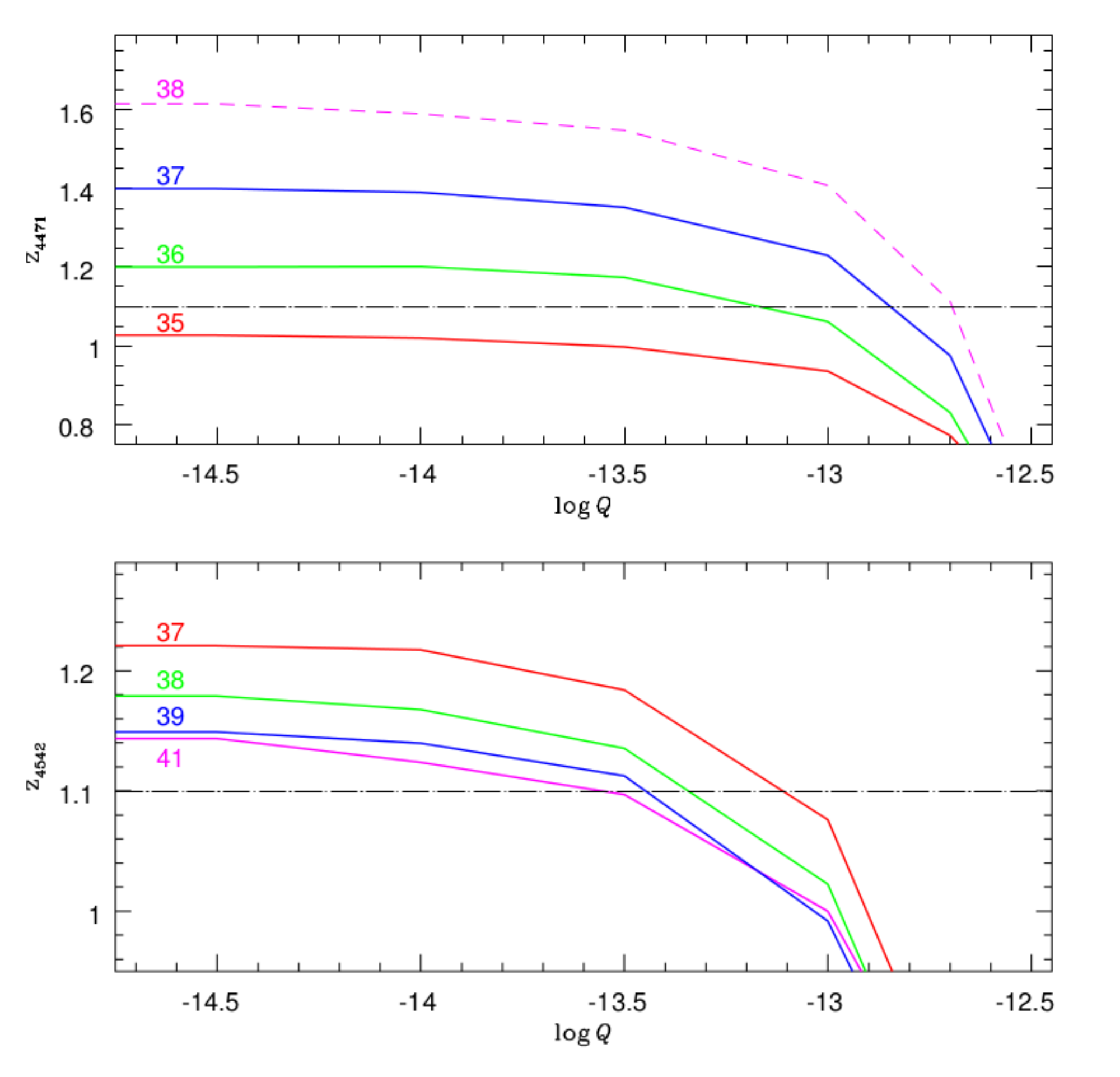}
\caption{{\sc fastwind} predictions of the behaviour of the EW ratios that
  define the z spectral characteristic, with the variation of the
  wind-strength parameter $Q$. Fixed values of the resolution ($R=2500$),
  surface   gravity (${\rm log}\,g=4.0$~dex), and projected rotational
  velocity ($v\,{\rm sin}\,i=100$~km\,s$^{-1}$) have been considered. The
  labels next to each color curve indicate the corresponding effective
  temperature expressed in kK. 
In the upper panel, the dashed line for 38 kK  indicates that this temperature
is likely out of the $z_{4471}$ relevant range.}  
\label{EW-vs-logQ}
\end{figure}

We first study how $z_{4542}$ and $z_{4471}$ behave with the
variation of the wind strength $Q$-parameter.
Fig.~\ref{EW-vs-logQ} shows these parameters as a function of log\,$Q$ for a set of
different values of the $T_{\rm eff}$, ranging from 35 to 38~kK for
$z_{4471}$ (top panel), and from 37 to 41~kK for $z_{4542}$ (bottom
panel). Resolution, surface gravity, and projected rotational velocity are
fixed for simplicity. 
In both panels, the dash-dot line at $z=1.10$ marks the value upon which the
star is classified as Vz.  

The first conclusion that arises from the top panel of Fig.~\ref{EW-vs-logQ}
is that O spectra with effective temperatures below a certain value (which is
close to 35~kK)  will never be classified as Vz.  
No matter how weak the stellar wind is, the $z$ parameter for the latest-type stars
never exceeds the adopted threshold.  
This conclusively justifies the decision to  exclude all the dwarfs with
spectral types later than O8.5 from the very beginning of the study
(Sec.~\ref{sel-crit}). 

Fig.~\ref{EW-vs-logQ} also shows that at higher temperatures 
the stars may show the Vz characteristic until the $Q$ parameter takes a 
certain value denoted as $Q_{\rm lim}$. 
This limit value of $Q$ would represent the strength of the wind required to
sufficiently fill in the He\,{\sc ii}~$\lambda$4686 absorption, so the
condition $z > 1.10$ is broken and, as already shown by SS14, is highly
dependent on $T_{\rm eff}$. In a solar-metallicity environment, $Q_{\rm lim}$
takes its largest value 
at 37~kK, considering either $z_{4542}$ or $z_{4471}$ (in the upper panel,
the curve for 38~kK is included for reference, but this temperature is very
likely out of the $z_{4471}$ relevant range).  
Therefore, the main-sequence stars whose $T_{\rm  eff}$ are close to 37~kK are
those that need the strongest stellar winds to no longer show the Vz
peculiarity in their spectra. This may explain the predominance of
intermediate spectral types in the histogram of Fig.~\ref{sp-d}.  

The  bottom panel of Fig.~\ref{EW-vs-logQ} 
shows that for very early objects with $T_{\rm eff}$ above $\approx39$~kK, the
Vz characteristic may vanish much before the development of a strong wind; in
other words, these stars could be classified as non-z even if they have
stellar winds with a strength comparable to other O\,Vz at later types.  This
naturally explains the observed drop in the number 
of O\,Vz stars for spectral types earlier than O6. 

In the present model configuration, the approximate values of ${\rm
  log}\,Q_{\rm lim}$ are: $-12.85$ for $T_{\rm eff}\approx37$~kK 
(or $-13.1$, if considered $z_{4542}$), 
$-13.20$ for $T_{\rm eff}\approx36$~kK, 
and lower than $-13.35$ for $T_{\rm  eff}$ of the order and above $38$~kK.  
We recall that, in this model,  
the resolution, surface gravity, and projected rotational velocity are fixed
for didactic purposes ($R=2500$, $\log g =4.0$, and
$v\sin i=100$). Although the overall behaviour of the $z$ parameter is similar, the
limit values of $Q$ can logically vary  when alternative configurations are
considered, i.e. for different values of $R$, $\log g$, and $v\sin i$, which
is a fortunate fact as the existence of Vz stars with  $T_{\rm eff}>39$~kK would
be otherwise very difficult to explain. 
Anyway, $Q_{\rm lim}$ for $T_{\rm eff}\approx37$~kK  always appears to be
significantly larger than those computed for other effective temperatures.  
As a conclusion we may say that, in a solar metallicity environment, the
O-type stars with effective temperatures close to that value are by far
the most ``favoured by nature'' to show the Vz characteristic in their 
spectra. This fact is in complete agreement with the observed statistics.  

We would like to emphasise that the goal of the very simple theoretical
analysis performed here is to contribute to understanding the observations, in
particular, the behaviour of the measured EWs, as well as the spectral-type
distribution observed for the Galactic O\,Vz and O\,V stars.  Thus, we show
that {\sc fastwind} model predictions can nicely account for some of the 
observed characteristics. These results  neither contradict nor confirm the
hypothesis of the extreme youth of the Vz stars, but only evidence, as already
done by SS14, the importance of considering several stellar
parameters related to different physical processes when interpretating the Vz
phenomenon.   

%**************************************************************************
% FIGURE 9
%**************************************************************************
%**************************************************************************

\section{Locations of the Vz stars}

When investigating a peculiar or new class of stars, it is important to also
consider their environments and companions, which may provide clues to their
nature.  The initial hypothesis regarding the Vz class was that it might be
younger than the typical class V, based upon the ``inverse Of'' interpretation
of the phenomenon as explained in the Introduction.  Thus, we discuss the
locations of the Vz stars in this section, with special attention to the most
extreme members of the class. 
In an attempt to quantify the definition of an  ``extreme''  Vz star, we will
consider the condition $z \geq 1.20$.  
Although somewhat arbitrary,  this limit is useful since 
ensures the membership to the Vz class, even in the cases  where measurement
errors are large and, at the same time, focus the discussion on  the most
particular 
objects, such as for example those shown in Figure~{\ref{extremas}, for which
  the  
depth of the He~{\sc ii}  $\lambda4686$ absorption relative to  the other He
lines is really striking.

Table~\ref{list-cumulos} lists the clusters and H~{\sc ii} regions containing 
Vz objects from our current sample of $197$ luminosity-class V, i.e dwarfs,
stars.  The 
number of z dwarfs within each cluster is quoted in the third column of the
table.  
The number of  non-z dwarfs is also provided for comparison in the last
column. Of course, these numbers are lower limits not only because of
observational incompleteness, but also due to the selection criteria for the
measurable sample (Sec.~\ref{sel-crit}). Column 4 details the names of the
z dwarfs in each cluster. The $39$ ``extreme'' members, as defined
previously, are marked in bold. As a comparison and control, 
Table~\ref{list-non-z} provides an analogous listing for the clusters that
contain normal, i.e. non-z,  dwarfs, but do
not include Vz objects. 

Table~\ref{list-cumulos} contains $52$ Vz stars in 15 clusters and/or H\,{\sc
  ii} regions that contain two or more Vz objects, all of which have
ages less than about 3 Ma insofar as is known; 
at greater cluster ages, H~{\sc ii} regions are generally
absent (Walborn 2010). The same Table also includes
23 single (to date) Vz stars in 23 young clusters and/or H~II regions, for a
grand total of $75$ found in such environments.  Only three Vz objects 
(HD~19265, HD~44811, and HD~97966),  
cannot be directly associated to a very young cluster.
The especially outstanding entries in Table~\ref{list-cumulos} 
are the Carina Nebula with a total of 19 Vz members, half of which are extreme
cases, and the extended Cygnus region with 12, distributed among various
subregions in both cases. 
The Vz content of the IC~1795/1805/1848 complex, and other clusters such
as NGC~6611 and NGC~1893, is also notable. These are all very young
objects, providing support for that interpretation of the Vz class. 

Besides the large absolute number of Vz stars in Carina and Cygnus, which
might be expected since those are the two best-represented
regions in GOSSS, the proportion of Vz to   normal   class V objects ($\sim 60$\,\%)
is evident.  Within the Carina Nebula, 
the difference between Trumpler~14 and Trumpler~16 is particularly
curious. While the former shows a majority of z objects, in the
latter there is a predominance of non-z  dwarfs. 
Some authors have suggested an extremely young age around 0.5 Ma for
Trumpler~14 (e.g. Penny et al. 1993; Rochau et al. 2011). 
On the other hand, although also very young, Trumpler~16 appears somewhat
older (e.g. Wolk et al. 2011) with evolved supergiants like the remarkable
Luminous Blue Variable $\eta$~Car.  

Among the clusters with no Vz objects (Table~\ref{list-non-z}), we emphasize
IC 2944, which contains at least 3 normal dwarfs (5, if the binaries
HD~101131~AB and HD~101436 are also taken into account), and shows a 
tenuous H\,{\sc ii} region. 
McSwain \& Gies (2005) evaluated the age of IC~2944 to be $6.6$~Ma.
Other salient cases are 
Havlen-Moffat~1, an evolved cluster containing Wolf-Rayet stars (Havlen \&
Moffat 1977), and also IC~1590 and NGC~6604, whose average ages are found to
be larger than 6.0~Ma (Kharchenko et al. 2013).
All the clusters from  Table~\ref{list-non-z} are on average older than those
containing a conspicuous population of O\,Vz objects, as suggested by
the different morphologies observed in the mid-infrared images from the 
{\em Wide-field Infrared Survey Explorer} ({\em WISE})
(Figure~\ref{wise}). 
The change in the morphology of a massive star forming region with the
evolutionary stage of the underlying massive stellar content is nicely
presented by Koenig et al. (2012). Using  the {\em WISE} images in the W3
channel ($12\,\mu$m), these authors show the correlation between the age of
the main 
exciting stellar cluster and the structure of the diffuse emission nebulosity
produced by excited polycyclic aromatic hydrocarbon molecules (PAH) and small
dust grains. Younger massive stellar clusters lack a cleared out
cavity around the powering O stars. Within a timescale of $3-5$ Ma, the H\,{\sc
  ii}  regions are cleared out and the observed PAH and dust emission is mostly
concentrated in peripheric ring structures with protuberances or pillars where
a new, triggered generation of stars is forming. This evolutive scenario of
massive star formation was previously depicted by Walborn \& Parker
(1992) (see also Walborn et al. 1999) and named "two-stage starbursts"
scenario.

Finally, in Figure~\ref{ha-all-sky} we present the spatial distribution of the
Galactic z (red circles) and non-z (cyan circles) O dwarfs,  overplotted on the
H$\alpha$ all-sky map produced by Finkbeiner (2004). 
The Vz objects, especially the most extreme, seem to show a higher
concentration toward the most active star forming regions. Of course,
normal dwarfs are also found within these regions but, as a
class, the latter appear more disseminated within and out of the Galactic disk. 
As a conclusion we can say that, while further astrophysical investigation is
needed, the extreme Vz objects may reasonably be hypothesized to
represent the youngest, post-embedded massive stars. 

%*****************************************************************************
% FIGURE 10
%*****************************************************************************
\begin{figure}
\includegraphics[angle=0,scale=0.4]{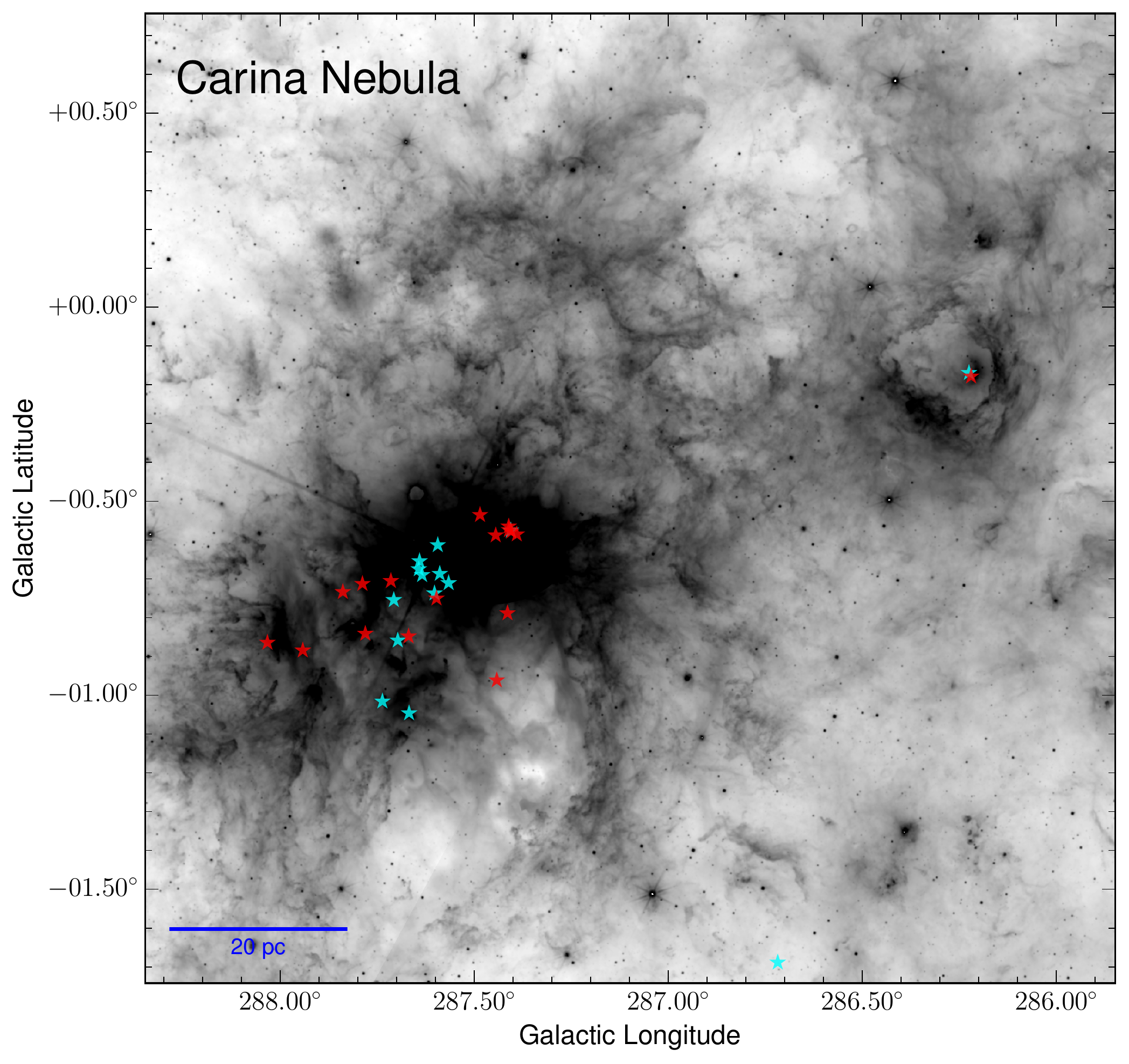}
\includegraphics[angle=0,scale=0.4]{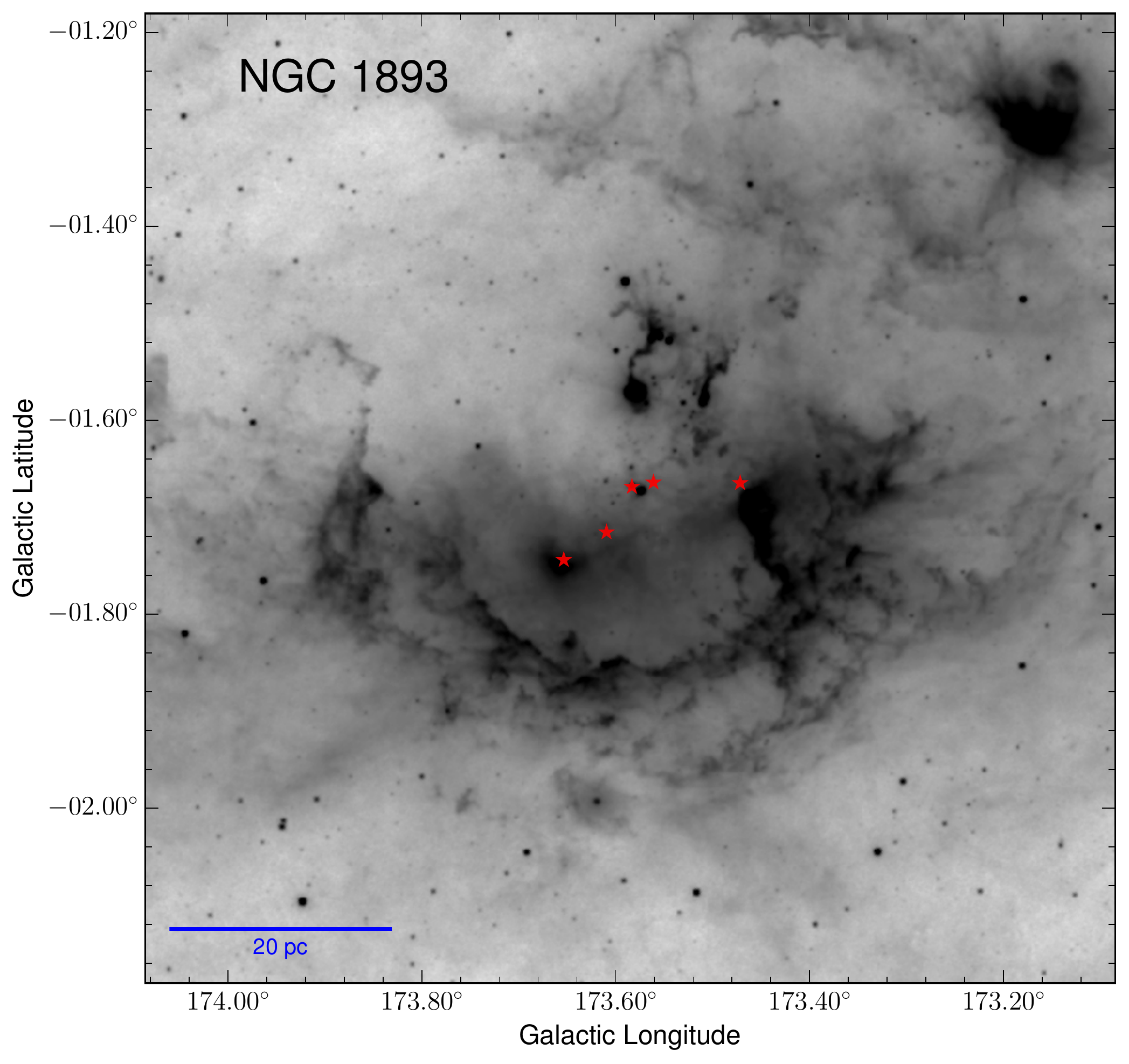}
\includegraphics[angle=0,scale=0.4]{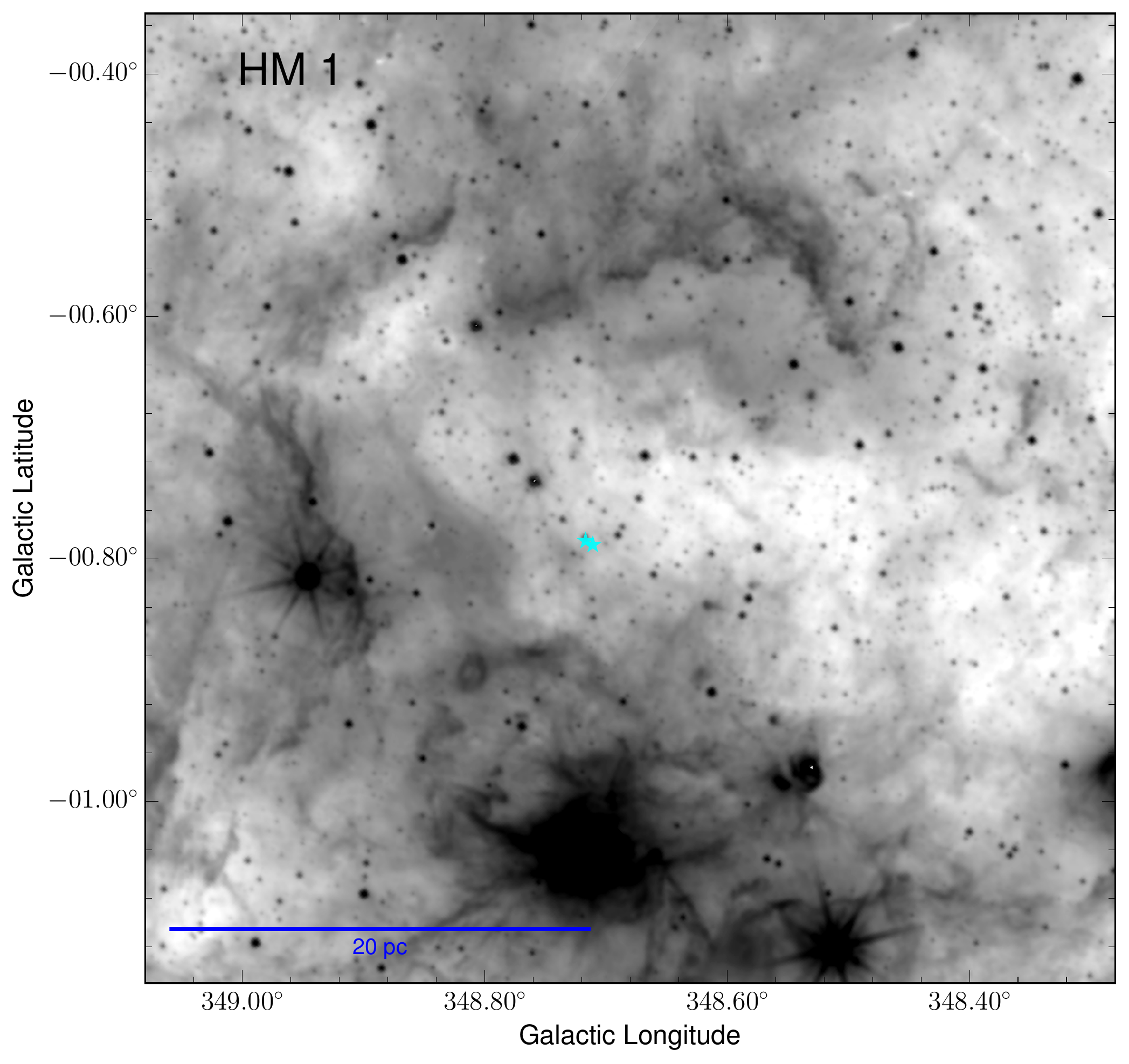}
\includegraphics[angle=0,scale=0.4]{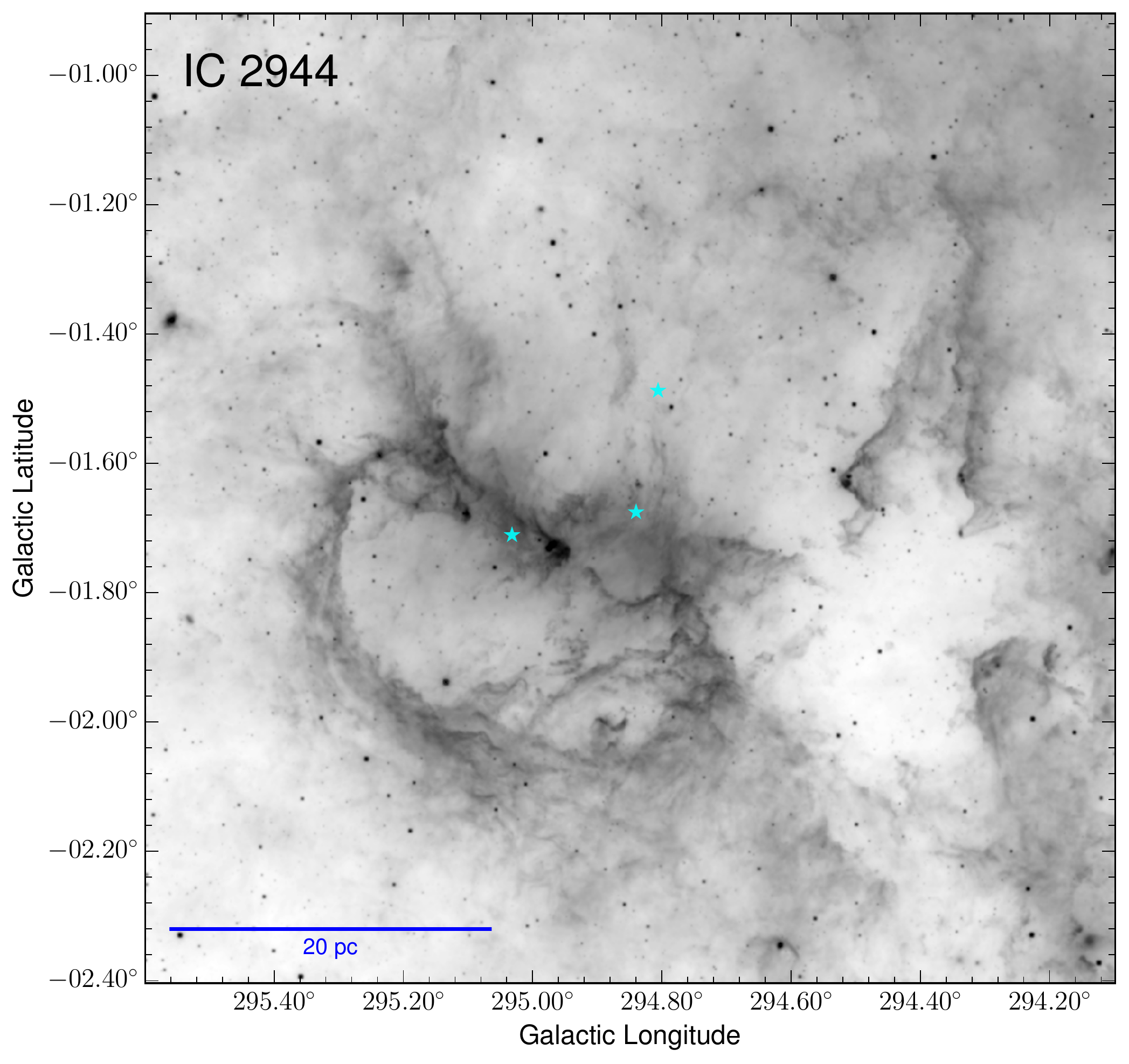}

\caption{Spatial distribution of Galactic O\,Vz (red stars) and O\,V (cyan
  stars) objects in selected H\,{\sc ii} regions. 
The background images correspond to the 12~$\mu$m W3 channel of the {\em Wide
Infrared Space Explorer} ({\em WISE}) represented in Galactic coordinates. 
The scale bar represents a length of 20~pc according to the distance
proposed for each region.
Top: Carina region (left) and NGC~1893 (right). Bottom: Havlen-Moffat 1 (left)
and IC 2944 (right).}  
\label{wise}
\end{figure}
%*****************************************************************************

%*****************************************************************************
% FIGURE 11
%*****************************************************************************
\begin{figure}
\includegraphics[angle=0,scale=2.5]{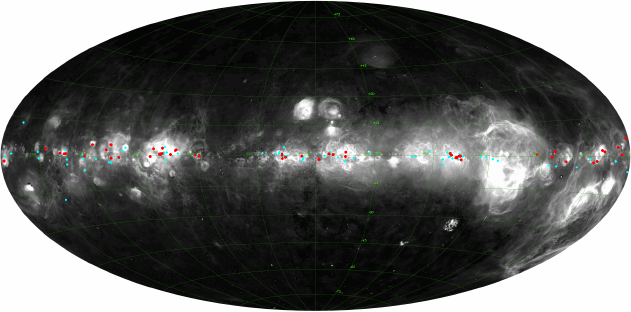}
\caption{Spatial distribution of Galactic O\,Vz (red circles) and O\,V (cyan
  circles) stars included in our sample. The background image is the H$\alpha$
  map in Aitoff projection produced by Finkbeiner (2004) and represented in
  Galactic  coordinates.}   
\label{ha-all-sky}
\end{figure}
%*****************************************************************************

\section{Summary and outlook}

Comprising the largest collection of O-star optical spectra ever assembled, The
Galactic O Star Spectroscopic Survey (GOSSS) provides a unique opportunity
to investigate in detail and systematically this important class of objects. 
The O stars belonging to the Vz luminosity subclass have been suggested to be
among the youngest optically observable massive objects, and thus their study
appears as key  
to the understanding of the early evolution of massive stars.

This work provided a new insight into the Vz spectral classification criteria. 
The convenience of incorporating a quantitative criterion based on the
equivalent widths (EWs) of the He\,{\sc i}~$\lambda$4471, He\,{\sc
  ii}~$\lambda$4542, and He\,{\sc ii}~$\lambda$4686 spectral lines was
shown. The EWs for a large sample of stars in the Galactic O Star Catalogue
(GOSC) were measured, allowing the recalibration of the Vz phenomenon. 
As a result, the threshold of the ``$z$  parameter'', defined as 
$z$~=~EW(He\,{\sc ii}~$\lambda$4686)/Max[EW(He\,{\sc i}~$\lambda$4471),EW(He\,{\sc  ii}~$\lambda$4542)], upon which a spectrum is classified as Vz, was
increased to 1.10. This change aims to avoid marginal cases, whose actual
membership to the Vz class is unclear and may be therefore contributing to
the recent controversy arised regarding the relation between the Vz
classification and the small age of the objects.
The new critical value for the $z$ parameter proposed in this work was adopted in
the 
third installment of the GOSSS project. This led to a complete revision of the
spectral types in GOSC and, more importantly, to the definition for the first
time, of O\,Vz standard stars.  

The population of O\,V and O\,Vz stars resulting from the new classification
criteria were comparatively analyzed. With $78$ members, the O\,Vz objects
represent $\sim 40\,\%$ of the dwarfs in the relevant spectral range of
O3-O8.5. However their distribution as a function of spectral type is 
not uniform but shows a marked concentration toward intermediate types. 
Based on a very simple theoretical analysis performed using synthetic spectra
from a grid of {\sc fastwind} models computed for a solar-metallicity value, we
showed that the former, as well as  
other observed characteristics in the spectral-type distribution of the
O\,Vz stars, are in perfect agreement with what is predicted by the models. 
This study confirmed the result from SS14 about the complexity, previously
unrecognized, of the Vz phenomenon and the necessity of taking into account many
processes for its correct interpretation.   

The locations, environment, and companions of the O\,Vz stars were analyzed,
with special attention to the most extreme members. Almost all of the
O\,Vz objects are associated with very young clusters, all of which have ages
less than 3~Ma insofar as is known. On the other hand, clusters with no Vz
objects are clearly older, as evidenced by the morphologies observed in the
mid-infrared WISE images, and the presence of more evolved stars. These facts
add considerable support for the interpretation of the Vz characteristic as a
signature of 
youth, although further astrophysical investigation is needed.

The results presented here represent the first steps toward our goal of
understanding the physical significance of the O\,Vz stars. We have defined
a robust sample of O\,Vz stars, many of them presumably single, suitable to
perform quantitative  
spectroscopic analyses in high-resolution which will allow us to derive their
stellar and wind parameters. 
Such a study, aimed to be presented in a forthcoming paper, will certainly
establish the actual evolutionary status of these interesting objects.

%********************************************************************
% TABLES 1,2,3,4,5
%********************************************************************
\begin{landscape}
\begin{deluxetable}{lllllcccc}
\tablewidth{0pt}
\tabletypesize{\footnotesize}
\tablecaption{O\,V stars in GOSSS with no evidence of binarity}
\startdata
\hline
  \multicolumn{1}{l}{Object} &
  \multicolumn{1}{c}{R.A.} &
  \multicolumn{1}{c}{Decl.} &
  \multicolumn{1}{l}{Sp. Type} &
  \multicolumn{1}{c}{Ref.} &
  \multicolumn{1}{c}{EW(He\,{\sc i}~4471)} &
  \multicolumn{1}{c}{EW(He\,{\sc ii}~4542)} &
  \multicolumn{1}{c}{EW(He\,{\sc ii}~4686)} &
  \multicolumn{1}{c}{$z$} \\

        & hh:mm:ss.ss & dd:am:as.s  &                      &      & [\AA] & [\AA] &  [\AA] &   \\
\hline
BD~+66~1675     & 00:02:10.29 & +67:24:32.2 &   O7.5\,Vz           & MA16 & 0.70 & 0.68 & 0.88 &  1.25  \\
Tyc~4026-00424-1 & 00:02:19.03 & +67:25:38.5 &   O7\,V((f))z       & MA16 & 0.55 & 0.66 & 0.84 &  1.27  \\
ALS~6351        & 00:48:12.55 & +62:59:24.8 &   O7\,Vz             & MA16 & 0.71 & 0.90 & 1.04 &  1.16  \\
HD~5005A        & 00:52:49.21 & +56:37:39.5 &   O4\,V((fc))        & S11a & 0.19 & 0.76 & 0.81 &  1.07  \\
HD~5005C        & 00:52:49.55 & +56:37:36.8 &   O8.5\,V(n)         & S11a & 0.85 & 0.81 & 0.76 &  0.89  \\
BD~+60~134      & 00:56:14.22 & +61:45:36.9 &   O5.5\,V(n)((f))    & MA16 & 0.50 & 0.91 & 1.00 &  1.10  \\
HD~5689         & 00:59:47.59 & +63:36:28.3 &   O7\,Vn ((f))       & MA16 & 0.70 & 0.92 & 0.71 &  0.78  \\
HD~12993        & 02:09:02.47 & +57:55:55.9 &   O6.5\,V((f))~Nstr  & S14  & 0.70 & 0.95 & 1.00 &  1.05  \\
BD~+61~411~A    & 02:26:34.39 & +62:00:42.4 &   O6.5\,V((f))z      & MA16 & 0.46 & 0.73 & 0.89 &  1.22  \\
BD~+60~501      & 02:32:36.27 & +61:28:25.6 &   O7\,V(n)((f))z     & S11a & 0.57 & 0.75 & 0.91 &  1.21  \\
HD~15629        & 02:33:20.59 & +61:31:18.2 &   O4.5\,V((fc))      & S11a & 0.20 & 0.66 & 0.52 &  0.78  \\
BD~+60~513      & 02:34:02.53 & +61:23:10.9 &   O7\,Vn             & MA16 & 0.61 & 0.82 & 0.75 &  0.91  \\
BD~+62~424      & 02:36:18.22 & +62:56:53.4 &   O6.5\,V(n)((f))    & S11a & 0.45 & 0.80 & 0.73 &  0.91  \\
HD~17505B       & 02:51:08.26 & +60:25:03.8 &   O8\,V              & S11a & 0.70 & 0.54 & 0.69 &  0.99  \\
BD~+60~586~A    & 02:54:10.67 & +60:39:03.6 &   O7\,Vz             & S11a & 0.68 & 0.67 & 0.83 &  1.23  \\
ALS~7833        & 03:59:07.49 & +57:14:11.7 &   O8\,Vz             & MA16 & 0.72 & 0.70 & 0.87 &  1.20  \\
BD~+50~886      & 04:03:20.74 & +51:18:52.5 &   O4\,V((fc))        & MA16 & 0.12 & 0.74 & 0.79 &  1.07  \\
BD~+52~805      & 04:18:35.64 & +52:51:54.3 &   O8\,V(n)           & MA16 & 0.80 & 0.77 & 0.86 &  1.07  \\
ALS~8294        & 05:22:39.69 & +33:22:18.2 &   O7\,V(n)z          & MA16 & 0.83 & 0.81 & 0.93 &  1.12  \\
HDE~242926      & 05:22:40.10 & +33:19:09.4 &   O7\,Vz             & S11  & 0.61 & 0.77 & 0.88 &  1.14  \\
BD~+33~1025~A   & 05:22:44.00 & +33:26:26.6 &   O7.5\,V(n)z        & S14  & 0.81 & 0.87 & 0.96 &  1.11  \\
HDE~242935~A    & 05:22:46.54 & +33:25:11.5 &   O6.5\,V((f))z      & S11a & 0.49 & 0.83 & 0.92 &  1.12  \\
HD~35619        & 05:27:36.15 & +34:45:19.0 &   O7.5\,V((f))       & MA16 & 0.73 & 0.64 & 0.71 &  0.97  \\
HD~36879        & 05:35:40.53 & +21:24:11.7 &   O7\,V(n)((f))      & MA16 & 0.60 & 0.76 & 0.72 &  0.95  \\
HD~41161        & 06:05:52.46 & +48:14:57.4 &   O8\,Vn             & S11  & 0.98 & 0.87 & 0.78 &  0.80  \\
HD~41997        & 06:08:55.82 & +15:42:18.2 &   O7.5\,Vn((f))      & S11  & 0.71 & 0.77 & 0.55 &  0.72  \\
HD~42088        & 06:09:39.57 & +20:29:15.5 &   O6\,V((f))z        & S11  & 0.63 & 0.81 & 1.06 &  1.30  \\
HD~44811        & 06:24:38.35 & +19:42:15.8 &   O7\,V(n)z          & S11  & 0.70 & 0.75 & 0.95 &  1.27  \\
ALS~19265       & 06:24:59.85 & +26:49:20.0 &   O4.5\,V((c))z      & MA16 & 0.31 & 1.10 & 1.24 &  1.13  \\
HDE~256725~A    & 06:25:01.30 & +19:50:56.1 &   O5\,V((fc))        & MA16 & 0.34 & 0.94 & 0.92 & 0.97  \\
HD~46056~A      & 06:31:20.86 & +04:50:03.9 &   O8\,Vn             & S11  & 0.80 & 0.69 & 0.75 &  0.94  \\
HD~46150        & 06:31:55.52 & +04:56:34.3 &   O5\,V((f))z        & S11  & 0.26 & 0.57 & 0.79 &  1.39  \\
HD~46573        & 06:32:09.31 & +04:49:24.7 &   O4\,V((f))         & S11  & 0.16 & 0.70 & 0.70 &  0.99  \\
HD~46485        & 06:33:50.96 & +04:31:31.6 &   O7\,V((f))n~var?   & MA16 & 0.80 & 0.78 & 0.82 &  1.03  \\
Tyc~0737-01170-1 & 06:36:29.00 & +10:49:20.7 &   O7\,Vz            & MA16 & 0.61 & 0.77 & 0.98 &  1.27  \\
HD~48279~A      & 06:42:40.55 & +01:42:58.2 &   O8.5\,V~Nstr~var?  & MA16 & 0.98 & 0.90 & 0.96 &  0.98  \\
ALS~207         & 07:09:55.20 & -18:30:07.9 &   O6.5\,V((f))       & MA16 & 0.57 & 0.91 & 0.98 &  1.08  \\
HD~57236        & 07:19:30.10 & -22:00:17.3 &   O8.5\,V            & S14  & 0.83 & 0.59 & 0.85 &  1.03  \\
ALS~458         & 07:30:01.27 & -19:08:35.0 &   O6.5\,V((f))z      & MA16 & 0.59 & 0.80 & 0.91 &  1.14  \\
BD~-15~1909     & 07:34:58.46 & -16:14:23.2 &   O6.5\,V((f))z      & MA16 & 0.61 & 0.81 & 0.93 &  1.15  \\
CPD~-26~2704    & 07:52:36.52 & -26:22:42.0 &   O7\,V(n)           & MA16 & 0.78 & 0.89 & 0.88 &  0.99  \\
HD~64568        & 07:53:38.21 & -26:14:02.6 &   O3\,V((f*))z       & S14  & 0.00 & 0.78 & 0.95 &  1.23  \\
CPD~-47~2962    & 08:57:51.66 & -47:45:43.9 &   O7\,V((f))         & MA16 & 0.64 & 0.95 & 0.86 &  0.91  \\
CPD~-49~2322    & 09:15:52.79 & -50:00:43.8 &   O7.5\,V((f))       & MA16 & 0.69 & 0.56 & 0.68 &  0.99  \\
HDE~298429      & 09:30:37.25 & -51:39:34.7 &   O8.5\,V            & S14  & 0.92 & 0.60 & 0.73 &  0.80  \\
2MASS J10224096-5930305 & 10:22:40.96 & -59:30:30.6 &   O7\,V((f))z & MA16 & 0.55 & 0.58 & 0.74 &  1.27  \\
2MASS J10224377-5930182 & 10:22:43.77 & -59:30:18.2 &   O8\,V(n)    & MA16 & 0.75 & 0.50 & 0.61 &  0.81  \\
HD~90273        & 10:23:44.45 & -57:38:31.5 &   ON7\,V((f))        & MA16 & 0.68 & 0.85 & 0.87 &  1.03  \\
HD~91837        & 10:34:49.51 & -60:11:14.1 &   O8.5\,Vn           & S14  & 0.84 & 0.62 & 0.72 &  0.86  \\
HD~92206+B      & 10:37:22.96 & -58:37:23.0 &   O6\,V((f))         & MA16 & 0.48 & 0.75 & 0.81 &  1.08  \\
HD~92504        & 10:39:36.87 & -57:27:40.7 &   O8.5\,V(n)         & S14  & 0.83 & 0.58 & 0.80 &  0.96  \\
HDE~305438      & 10:42:43.77 & -59:54:16.5 &   O8\,Vz             & S14  & 0.73 & 0.89 & 0.98 &  1.10  \\
HDE~303316~A    & 10:43:11.18 & -59:44:21.0 &   O7\,V((f))z        & S14  & 0.52 & 0.64 & 0.90 &  1.42  \\
CPD~-58~2611    & 10:43:46.69 & -59:32:54.8 &   O6\,V((f))z        & S14  & 0.45 & 0.71 & 0.92 &  1.29  \\
HD~93128        & 10:43:54.37 & -59:32:57.4 &   O3.5\,V((fc))z     & S14  & 0.08 & 0.68 & 0.79 &  1.16  \\
Trumpler~14-9   & 10:43:55.35 & -59:32:48.6 &   O8.5\,V            & S14  & 0.85 & 0.67 & 0.87 &  1.03  \\
HD~93129~B      & 10:43:57.64 & -59:32:53.5 &   O3.5\,V((f))z      & MA16 & 0.13 & 0.74 & 0.83 &  1.12  \\
CPD~-58~2620    & 10:43:59.92 & -59:32:25.4 &   O7\,V((f))z        & S14  & 0.59 & 0.66 & 0.96 &  1.46  \\
HD~93204        & 10:44:32.34 & -59:44:31.0 &   O5.5\,V((f))       & S14  & 0.33 & 0.81 & 0.65 &  0.80  \\
HDE~303311      & 10:44:37.46 & -59:32:55.4 &   O6\,V((f))z        & S14  & 0.41 & 0.78 & 1.02 &  1.31  \\
HDE~305524      & 10:44:45.24 & -59:54:41.6 &   O6.5\,Vn((f))z     & S14  & 0.57 & 0.67 & 0.83 &  1.23  \\
CPD~-59~2610    & 10:44:54.71 & -59:56:01.9 &   O8.5\,V            & S14  & 0.81 & 0.78 & 0.84 &  1.04  \\
$[$ARV2008$]$~206 & 10:45:22.28 & -59:50:47.1 & O6\,V((f))         & S14  & 0.40 & 0.68 & 0.66 &  0.98 \\
HDE~305532      & 10:45:34.07 & -59:57:26.7 &   O6.5\,V((f))z      & S14  & 0.49 & 0.72 & 0.95 &  1.32  \\
%HDE~305525      & 10:46:05.70 & -59:50:49.4 &   O5.5V(n)((f))z     & S14  & 0.00 & 0.78 & 0.87 &  1.12  \\
CPD~-59~2673    & 10:46:22.46 & -59:53:20.5 &   O5.5V(n)((f))z     & S14  & 0.40 & 0.76 & 0.92 &  1.21  \\
HDE~305539      & 10:46:33.07 & -60:04:12.6 &   O8\,Vz             & S14  & 0.79 & 0.73 & 0.89 &  1.13  \\
HDE~305612      & 10:47:16.42 & -60:05:40.0 &   O8\,V(n)z          & S14  & 0.75 & 0.68 & 0.83 &  1.11  \\
HD~96715        & 11:07:32.82 & -59:57:48.7 &   O4\,V((f))z        & S14  & 0.21 & 0.67 & 0.94 &  1.40  \\
HD~97848        & 11:14:31.90 & -59:01:28.8 &   O8\,V              & S14  & 0.82 & 0.67 & 0.76 &  0.92  \\
NGC~3603~HST-51 & 11:15:07.50 & -61:15:46.3 &   O5.5\,V(n)         & MA16 & 0.22 & 0.64 & 0.52 &  0.81  \\
NGC~3603~MTT~25 & 11:15:11.32 & -61:15:55.6 &   O5\,V(n)           & MA16 & 0.23 & 0.66 & 0.49 &  0.74  \\
HD~97966        & 11:15:11.78 & -59:24:58.3 &   O7\,V((f))z        & MA16 & 0.54 & 0.61 & 0.86 &  1.40  \\
HD~99546        & 11:26:36.90 & -59:26:13.6 &   O7.5\,V((f))~Nstr  & MA16 & 0.72 & 0.82 & 0.78 &  0.95  \\
HD~101223       & 11:38:22.77 & -63:12:02.8 &   O8\,V              & S14  & 0.78 & 0.51 & 0.72 &  0.92  \\
HD~110360       & 12:42:12.70 & -60:39:08.7 &   ON7\,V             & MA16 & 0.74 & 1.02 & 1.06 &  1.04  \\
CPD~-61~3973    & 13:45:21.10 & -62:25:35.4 &   O7.5\,V((f))       & MA16 & 0.76 & 0.79 & 0.84 &  1.06  \\
HD~122313       & 14:03:12.99 & -62:15:38.6 &   O8.5\,V            & MA16 & 0.91 & 0.56 & 0.85 &  0.93  \\
HD~144647       & 16:09:16.20 & -49:36:21.8 &   O8.5\,V(n)         & MA16 & 0.88 & 0.63 & 0.81 &  0.91  \\
HDE~329100~A    & 16:54:42.30 & -45:15:14.8 &   O8\,V(n)           & MA16 & 0.79 & 0.62 & 0.64 &  0.81  \\
HDE~326775      & 17:05:31.31 & -41:31:20.1 &   O6.5V(n)((f))z     & MA16 & 0.48 & 0.69 & 0.80 &  1.16  \\
ALS~18770       & 17:19:00.80 & -38:49:23.1 &   O7\,V((f))         & MA16 & 0.55 & 0.75 & 0.76 &  1.01  \\
ALS~18768       & 17:19:01.05 & -38:48:58.9 &   O8.5\,V            & S14  & 1.03 & 0.52 & 1.00 &  0.97  \\
Pismis~24-15    & 17:24:28.95 & -34:14:50.7 &   O7.5\,Vz           & MA16 & 0.83 & 0.66 & 0.93 &  1.12  \\
ALS~17696       & 17:24:42.33 & -34:13:21.4 &   O7.5:\,V           & S14  & 0.76 & 1.13 & 0.87 &  0.77  \\
Pismis~24-2     & 17:24:43.31 & -34:12:44.2 &   O5\,V((f))         & S14  & 0.31 & 0.74 & 0.80 &  1.08  \\
ALS~16052       & 17:24:45.78 & -34:09:39.9 &   O6\,V((f))z        & S14  & 0.48 & 0.65 & 0.87 &  1.34  \\
HD~164492~A     & 18:02:23.55 & -23:01:51.1 &   O7.5\,Vz           & S14  & 0.66 & 0.82 & 1.05 &  1.27  \\
ALS~4626        & 18:04:17.88 & -13:06:13.7 &   ON6\,V((f))        & MA16 & 0.50 & 1.13 & 1.11 &  0.98  \\
HD~167633       & 18:16:49.66 & -16:31:04.3 &   O6.5\,V((f))       & S11  & 0.48 & 0.86 & 0.80 &  0.93  \\
ALS~4880        & 18:17:33.67 & -12:05:42.8 &   O6\,V((f))         & MA16 & 0.47 & 0.72 & 0.76 &  1.05  \\
ALS~15360       & 18:18:37.48 & -13:43:39.2 &   O7\,V((f))z        & MA16 & 0.50 & 0.80 & 0.90 &  1.13  \\
HD~168461       & 18:20:17.18 & -12:10:19.2 &   O7.5\,V((f))Nstr   & MA16 & 0.72 & 0.84 & 0.77 &  0.92  \\
HD~168504       & 18:20:34.10 & -13:57:15.7 &   O7.5\,V(n)z        & MA16 & 0.71 & 0.75 & 0.88 &  1.18  \\
ALS~19618       & 18:20:34.50 & -16:10:11.7 &   O4\,V(n)((fc))     & MA16 & 0.18 & 0.78 & 0.75 &  0.96  \\
BD~-14~5014     & 18:22:22.31 & -14:37:08.5 &   O7.5\,V(n)((f))    & MA16 & 0.55 & 0.50 & 0.58 &  1.06  \\
BD~-10~4682     & 18:24:20.65 & -10:48:34.3 &   O7\,Vn((f))        & MA16 & 0.66 & 0.90 & 0.82 &  0.90  \\
BD~-14~5040     & 18:25:38.90 & -14:45:05.7 &   O5.5\,V(n)((f))    & MA16 & 0.38 & 0.77 & 0.83 &  1.07  \\
BD~-04~4503     & 18:35:32.54 & -04:47:55.4 &   O7\,V              & MA16 & 0.79 & 0.95 & 0.92 &  0.97  \\
HDE~344758      & 19:41:52.72 & +24:20:51.1 &   O8.5\,V(n)((f))    & MA16 & 0.83 & 0.69 & 0.70 &  0.84  \\
HDE~344777      & 19:42:11.47 & +23:26:00.5 &   O7.5\,Vz           & MA16 & 0.69 & 0.70 & 0.79 &  1.12  \\
HDE~344784~A    & 19:43:10.97 & +23:17:45.4 &   O6.5\,V((f))z      & MA16 & 0.48 & 0.58 & 0.69 &  1.18  \\
HDE~338916      & 19:45:42.12 & +25:21:16.4 &   O7.5\,Vz           & MA16 & 0.73 & 0.84 & 0.95 &  1.13  \\
HDE~227018      & 19:59:49.10 & +35:18:33.5 &   O6.5\,V((f))z      & MA16 & 0.47 & 0.69 & 0.87 &  1.26  \\
HDE~227245      & 20:02:21.71 & +35:40:29.8 &   O7\,V((f))z        & MA16 & 0.56 & 0.66 & 0.87 &  1.32  \\
HDE~227465      & 20:04:27.22 & +33:42:18.4 &   O7\,V((f))         & MA16 & 0.66 & 0.68 & 0.64 &  0.94  \\
HD~191978       & 20:10:58.28 & +41:21:09.9 &   O8\,V              & MA16 & 0.80 & 0.54 & 0.82 &  1.03  \\
HDE~228759      & 20:17:07.54 & +41:57:26.5 &   O6.5V(n)((f))z     & MA16 & 0.53 & 0.67 & 0.80 &  1.20  \\
ALS~18707       & 20:17:41.93 & +36:45:25.6 &   O6.5\,V((f))z      & MA16 & 0.44 & 0.70 & 0.79 &  1.13  \\
HDE~228841      & 20:18:29.69 & +38:52:39.8 &   O6.5\,Vn((f))      & S11  & 0.60 & 0.89 & 0.83 &  0.93  \\
HD~193595       & 20:19:31.33 & +39:03:26.2 &   O7\,V((f))         & MA16 & 0.66 & 0.92 & 0.99 &  1.08  \\
HDE~229202      & 20:23:22.84 & +40:09:22.5 &   O7.5\,V(n)((f))    & MA16 & 0.65 & 0.64 & 0.66 &  1.01  \\
ALS~11355       & 20:27:17.57 & +39:44:32.6 &   O8\,V(n)((f))      & MA16 & 0.75 & 0.59 & 0.71 &  0.95  \\
BD~+40~4179     & 20:27:43.62 & +40:35:43.5 &   O8\,Vz             & MA16 & 0.70 & 0.72 & 0.81 &  1.12  \\
2MASS J20315961+4114504 & 20:31:59.61 & +41:14:50.5 &   O7.5\,Vz   & MA16 & 0.62 & 0.81 & 0.96 &  1.18  \\
Cyg~OB2-6       & 20:32:45.45 & +41:25:37.5 &   O8.5\,V(n)         & MA16 & 0.79 & 0.61 & 0.71 &  0.91  \\
ALS~15111       & 20:32:59.13 & +41:24:25.0 &   O8\,V              & MA16 & 0.82 & 0.59 & 0.83 &  1.01 \\
Cyg~OB2-22~B    & 20:33:08.83 & +41:13:17.4 &   O6\,V((f))         & S11  & 0.43 & 0.66 & 0.56 &  0.85  \\
Cyg~OB2-8~D     & 20:33:16.33 & +41:19:02.0 &   O8.5\,V(n)         & S14  & 0.75 & 0.46 & 0.76 &  1.02  \\
Cyg~OB2-24      & 20:33:17.48 & +41:17:09.3 &   O8\,V(n)           & S11  & 0.71 & 0.74 & 0.68 &  0.92  \\
Cyg~OB2-25~A    & 20:33:25.54 & +41:33:26.7 &   O8\,Vz             & MA16 & 0.69 & 0.59 & 0.82 &  1.19  \\
ALS~15134       & 20:33:26.76 & +41:10:59.5 &   O8\,Vz             & MA16 & 0.76 & 0.67 & 0.87 &  1.15  \\
BD~+45~3216~A   & 20:33:50.37 & +45:39:40.9 &   O5\,V((f))z        & MA16 & 0.38 & 0.86 & 1.08 &  1.26  \\
BD~+36~4145     & 20:36:18.21 & +37:25:02.8 &   O8.5\,V(n)         & MA16 & 0.78 & 0.71 & 0.72 &  0.92  \\
ALS~12050       & 21:55:15.29 & +57:39:45.7 &   O5\,V((f))         & MA16 & 0.25 & 0.74 & 0.64 &  0.86  \\
BD~+55~2722~A   & 22:18:58.63 & +56:07:23.5 &   O8\,Vz             & MA16 & 0.73 & 0.72 & 0.84 &  1.14  \\
ALS~12370       & 22:23:17.42 & +55:38:02.3 &   O6.5\,Vnn((f))     & MA16 & 0.61 & 0.92 & 0.89 &  0.97  \\
HD~213023~A     & 22:26:52.36 & +63:43:04.9 &   O7.5\,Vz           & MA16 & 0.69 & 0.72 & 0.86 &  1.20  \\
ALS~12619       & 22:47:50.60 & +58:05:12.4 &   O7\,V((f))z        & MA16 & 0.71 & 0.81 & 1.10 &  1.36  \\
HD~216532       & 22:52:30.56 & +62:26:25.9 &   O8.5\,V(n)         & S11  & 0.93 & 0.69 & 0.77 &  0.83  \\
BD~+55~2840     & 22:55:08.49 & +56:22:58.9 &   O7.5\,V(n)         & MA16 & 0.72 & 0.79 & 0.76 &  0.96  \\
HD~217086       & 22:56:47.19 & +62:43:37.6 &   O7\,Vnn((f))z      & S14  & 0.61 & 0.69 & 0.83 &  1.20  \\
BD~+60~2635     & 23:53:05.21 & +60:54:44.6 &   O6\,V((f))         & MA16 & 0.48 & 0.91 & 0.95 &  1.04  \\
\label{list-singles}
\enddata
\end{deluxetable}

\end{landscape}

\begin{landscape}
\begin{deluxetable}{lllllcccc}
\tablewidth{0pt}
\tabletypesize{\footnotesize}
\tablecaption{O binaries with single-lined profiles at the GOSSS resolution}
\startdata
\hline
  \multicolumn{1}{l}{Object} &
  \multicolumn{1}{c}{R.A.} &
  \multicolumn{1}{c}{Decl.} &
  \multicolumn{1}{l}{Sp. Type} &
  \multicolumn{1}{c}{Ref.} &
  \multicolumn{1}{c}{EW(He\,{\sc i}~4471)} &
  \multicolumn{1}{c}{EW(He\,{\sc ii}~4542)} &
  \multicolumn{1}{c}{EW(He\,{\sc ii}~4686)} &
  \multicolumn{1}{c}{$z$} \\
        & hh:mm:ss.ss & dd:am:as.s  &                      &      & [\AA] & [\AA] &  [\AA] &   \\
\hline
V747~Cep        & 00:01:46.87 & +67:30:25.1 & O5.5\,V(n)((f))   & MA16 & 0.39 & 0.70 & 0.66 & 0.94   \\
HD~14633~AaAb   & 02:22:54.29 & +41:28:47.7 & ON8.5\,V          & S11  & 1.01 & 0.78 & 0.72 & 0.71  \\
HD~17520~A      & 02:51:14.43 & +60:23:10.0 & O8\,V             & MA16 & 0.80 & 0.75 & 0.68 & 0.85  \\
HDE~242908      & 05:22:29.30 & +33:30:50.4 & O4.5\,V(n)((fc))z & S14  & 0.28 & 0.74 & 0.94 & 1.26  \\
HD~46149        & 06:31:52.53 & +05:01:59.2 & O8.5\,V           & S11  & ...  & ...  & ...  & ...   \\
HD~46573        & 06:34:23.57 & +02:32:02.9 & O7\,V((f))z       & S11  & 0.72 & 0.77 & 0.88 & 1.15  \\
15~Mon~AaAb     & 06:40:58.66 & +09:53:44.7 & O7\,V((f))z~var   & S14  & 0.70 & 0.67 & 0.78 & 1.12  \\
ALS~85          & 06:45:48.84 & -07:18:46.4 & O7.5\,V           & MA16 & 0.69 & 0.86 & 0.89 & 1.04  \\
HD~53975        & 07:06:35.96 & -12:23:38.2 & O7.5\,Vz          & S11  & 0.75 & 0.70 & 0.83 & 1.11  \\
HD~54662~AB     & 07:09:20.25 & -10:20:47.6 & O7\,Vz~var?       & S14  & ...  & ...  & ...  & ...   \\
V467~Vel        & 08:43:49.81 & -46:07:08.8 & O6.5\,V(n)((f))   & MA16 & 0.51 & 0.62 & 0.52 & 0.83  \\
HD~91572        & 10:33:12.27 & -58:10:13.6 & O6.5\,V((f))z     & S14  & 0.50 & 0.61 & 0.74 & 1.22  \\
HD~91824        & 10:34:46.63 & -58:09:22.0 & O7\,V((f))z       & S14  & 0.60 & 0.78 & 0.89 & 1.14  \\
HD~92206~A      & 10:37:22.28 & -58:37:22.8 & O6\,V((f))z       & S14  & ...  & ...  & ...  & ...   \\
HD~93146~A      & 10:44:00.16 & -60:05:09.9 & O7\,V((f))        & MA16 & 0.68 & 0.79 & 0.78 & 0.98  \\
HD~93222~AB     & 10:44:36.25 & -60:05:28.9 & O7\,V((f))        & MA16 & 0.62 & 0.66 & 0.72 & 1.09  \\
CPD~-59~2600    & 10:44:41.79 & -59:46:56.4 & O6\,V((f))        & S14  & 0.40 & 0.69 & 0.65 & 0.94  \\
CPD~-59~2626~AB & 10:45:05.79 & -59:45:19.6 & O7.5\,V(n)        & MA16 & 0.73 & 0.78 & 0.72 & 0.92  \\
HDE~303308~AB   & 10:45:05.92 & -59:40:05.9 & O4.5\,V((fc))     & S14  & 0.23 & 0.64 & 0.55 & 0.86  \\
CPD~-59~2641    & 10:45:16.52 & -59:43:37.0 & O6\,V((fc))       & S14  & 0.33 & 0.63 & 0.55 & 0.87  \\
HD~101191       & 11:38:12.17 & -63:23:26.8 & O8\,V             & S14  & 0.74 & 0.50 & 0.76 & 1.03  \\
HD~101413~AB    & 11:39:45.84 & -63:28:40.1 & O8\,V             & S14  & 0.76 & 0.63 & 0.65 & 0.85  \\
HD~101436       & 11:39:49.96 & -63:28:43.6 & O6.5\,V((f))      & S14  & ...  & ...  & ...  & ...   \\
HD~123590~AB    & 14:10:43.97 & -62:28:44.4 & O8\,V             & S14  &  ... & ...  & ...  & ...   \\
HD~145217       & 16:12:00.30 & -50:18:20.5 & O8\,V             & MA16 & ...  & ...  & ...  & ...   \\
HD~150135~AaAb  & 16:41:19.45 & -48:45:47.5 & O6.5\,V((f))z     & S14  & ...  & ...  & ...  & ...   \\
HD~152590       & 16:56:05.22 & -40:20:57.6 & O7.5\,Vz          & S14  & 0.68 & 0.71 & 0.90 & 1.27  \\
HD~152623~AaAbB & 16:56:15.03 & -40:39:35.8 & O7\,V(n)((f))     & MA16 & ...  & ...  & ...  & ...   \\
HD~155913       & 17:16:26.34 & -42:40:04.1 & O4.5\,Vn((f))     & S14  & 0.31 & 0.86 & 0.75 & 0.88  \\
HDE~319699      & 17:19:30.42 & -35:42:36.1 & O5\,V((fc))       & S14  & 0.25 & 0.73 & 0.57 & 0.79  \\
HDE~319703~BaBb & 17:19:45.05 & -36:05:47.0 & O6\,V((f))z       & S14  & 0.47 & 0.60 & 0.71 & 1.19  \\
ALS~19693       & 17:25:29.17 & -34:25:15.7 & O6\,Vn((f))       & MA16 & 0.38 & 0.76 & 0.74 & 0.97  \\
%ALS~19692       & 17:25:34.23 & -34:23:11.7 & O5.5\,IV-V(f)     & MA16 & 0.31 & 0.61 & 0.56 & 0.92 \\
HD~164536       & 18:02:38.62 & -24:15:19.4 & O7.5\,V(n)        & MA16 & 0.75 & 0.71 & 0.81 & 1.07  \\
9~Sgr~AB        & 18:03:52.45 & -24:21:38.6 & O4\,V((f))        & MA16 & ...  & ...  & ...  & ...   \\
HD~165246       & 18:06:04.68 & -24:11:43.9 & O8\,V(n)          & S14  & 0.81 & 0.59 & 0.75 & 0.93  \\
HD~168075       & 18:18:36.04 & -13:47:36.5 & O7\,V((f))        & MA16 & ...  & ...  & ...  & ...   \\
HD~168137~AaAb  & 18:18:56.19 & -13:48:31.1 & O8\,Vz            & MA16 & 0.72 & 0.72 & 0.88 & 1.22  \\
BD~-16~4826     & 18:21:02.23 & -16:01:00.9 & O5.5\,V((f))z     & MA16 & 0.42 & 0.69 & 0.84 & 1.21 \\
V479~Sct        & 18:26:15.05 & -14:50:54.3 & ON6\,V((f))z      & MA16 & 0.51 & 0.94 & 1.05 & 1.12  \\
Cyg~OB2-17      & 20:32:50.01 & +41:23:44.7 & O8\,V             & MA16 & 0.77 & 0.80 & 0.80 & 1.00  \\
ALS~15115       & 20:33:18.05 & +41:21:36.9 & O8\,V             & MA16 & ...  & ...  & ...  & ...   \\
%ALS~15108~AB    & 20:33:23.47 & +41:09:13.1 & O5.5\,V((f))      & MA16 & ...  & ...  & ...  & ...  \\
Cyg~OB2-29      & 20:34:13.50 & +41:35:03.0 & O7.5\,V(n)((f))z  & MA16 & 0.62 & 0.62 & 0.79 & 1.27  \\
ALS~15114       & 20:34:29.60 & +41:31:45.4 & O7.5\,V(n)((f))   & MA16 & ...  & ...  & ...  & ...   \\
HD~199579       & 20:56:34.78 & +44:55:29.0 & O6.5\,V((f))z     & S11  & 0.57 & 0.60  & 0.85 & 1.40  \\
BD~+62~2078     & 22:25:33.58 & +63:25:02.6 & O7\,V((f))z       & MA16 & 0.62 & 0.82 & 0.92 & 1.13 \\
\label{list-sb1}
\enddata
\end{deluxetable}

\end{landscape}

\begin{landscape}
\begin{deluxetable}{lllclcccc}
\tablewidth{0pt}
\tabletypesize{\scriptsize}
\tablecaption{Double-lined binary stars in GOSSS}
\startdata
\hline
  \multicolumn{1}{l}{Object} &
  \multicolumn{1}{c}{R.A.} &
  \multicolumn{1}{c}{Decl.} &
  \multicolumn{1}{c}{Sp. Classif.} &
%  \multicolumn{1}{l}{Second.} &
  \multicolumn{1}{c}{Ref.} &
  \multicolumn{1}{c}{EW(He\,{\sc i}~4471)} &
  \multicolumn{1}{c}{EW(He\,{\sc ii}~4542)} &
  \multicolumn{1}{c}{EW(He\,{\sc ii}~4686)} &
  \multicolumn{1}{c}{$z$} \\
                & hh:mm:ss.ss & dd:am:as.s    &                &     & [\AA]& [\AA]& [\AA]&   \\
                &             &               & Prim. + Second.&     &Prim./Second.&Prim./Second. &Prim./Second. &Prim./Second.\\
\hline
ALS~6967        & 02:12:29.97 & +59:54:04.1   & O8\,V           + B0:\,V      & MA16 & 0.63 & 0.38 & 0.52 & 0.83  \\
BD~+60~497      & 02:31:57.09 & +61:36:43.9   & O6.5\,V((f))    + O8/B0\,V    & S11 & ...  & ...  & ...  & ...   \\
HD~18326        & 02:59:23.17 & +60:33:59.5   & O6.5\,V((f))z   + O9/B0\,V:   & S14  & ...  & ...  & ...  & ...   \\
MY~Cam          & 03:59:18.29 & +57:14:13.7   & O5.5\,V(n)      + O6.5\,V(n)  & MA16 & 0.23/0.29 & 0.78/0.36 & 0.51/0.38 & 0.65/1.06 \\
ALS~8272        & 05:20:00.63 & +38:54:43.5   & O7\,V((f))      + B0 III-V    & MA16 & 0.44 & 0.58 & 0.53 & 0.92  \\
HD~48099        & 06:41:59.23 & +06:20:43.5   & O5.5\,V((f))z   + O9\,V       & MA16 & ...  & ...  & ...  & ...   \\
HD~64315~AB     & 07:52:20.28 & -26:25:46.7   & O5.5\,V         + O7\,V       & MA16 & 0.34/0.37 & 0.61/0.41 & 0.63/0.35 & 1.03/0.84 \\
HD~92206~C      & 10:37:18.63 & -58:37:41.7   & O8\,V(n)z       + B0:\,V      & MA16 & 0.46 & 0.45 & 0.53 & 1.17  \\
ALS~15204       & 10:43:41.24 & -59:35:48.2   & O7.5\,Vz        + O9:\,V      & MA16 & ...  & ...  & ...  & ...   \\
HD~93161~A$^*$& 10:44:08.84 & -59:34:34.5   & O7.5\,V         + O9\,V       & S14  & 0.42 & 0.60 & 0.67 & 1.11  \\
HD~93205        & 10:44:33.74 & -59:44:15.5   & O3.5\,V((f))    + O8\,V       & S14  & ...  & ...  & ...  & ...   \\
CPD~-59~2591    & 10:44:36.69 & -59:47:29.6   & O8\,Vz          + B0.5:\,V:   & MA16 & 0.58 & 0.58 & 0.69 & 1.19  \\
V572~Car        & 10:44:47.31 & -59:43:53.2   & O7.5\,V(n)      + B0\,V(n)    & MA16 & 0.63 & 0.44 & 0.57 & 0.91  \\
HD~93343        & 10:45:12.22 & -59:45:00.4   & O8\,V           + sec         & MA16 & ...  & ...  & ...  & ...   \\
CPD~-59~2635    & 10:45:12.72 & -59:44:46.2   & O8\,V(n)        + O9.5\,V     & S14  & 0.57 & 0.55 & 0.42 & 0.72  \\
CPD~-59~2636~AB & 10:45:12.87 & -59:44:19.2   & O8\,V           + O8\,V       & S14  & ...  & ...  & ...  & ...   \\
V662~Car        & 10:45:36.32 & -59:48:23.2   & O5\,V(n)z       + B0:\,V      & MA16 & 0.24 & 0.77 & 0.85 & 1.11  \\
HDE~305525      & 10:46:05.70 & -59:50:49.4   & O5.5V(n)((f))z + sec          & S14  & ...  & ...  & ...  & ...  \\
ALS~18551$^*$  & 10:58:17.68 & -61:12:03.5   & O4.5\,V(n)z     + O4.5\,V(n)z & MA16 & 0.11/0.07 & 0.47/0.45 & 0.44/0.38 & 0.93/0.85 \\
2MASS~J10583238-6110565 & 10:58:32.39 & -61:10:56.5   & O5\,V((f))      + O7\,V((f))  & MA16 & ...  & ...  & ...  & ...   \\
HD~97484      & 11:12:04.50 & -61:05:42.9   & O7.5\,V((f))    + O7.5\,V((f))& MA16 & 0.37/0.34 & 0.46/0.33 & 0.34/0.27 & 0.74/0.79  \\
HD~100213     & 11:31:10.93 & -65:44:32.1   & O8\,V(n)        + B0\,V(n)    & MA16 & 0.54 & 0.45 & 0.53 & 0.98  \\
HD~101131~AB    & 11:37:48.44 & -63:19:23.5   & O5.5\,V((f))    + O8:\,V      & S14  & ...  & ...  & ...  & ...   \\
Tyc~7370-00460-1 & 17:18:15.40 & -34:00:05.9  & O6\,V((f))      + O8\,V       & MA16 & 0.33/0.22 & 0.41/0.37 & 0.43/0.38 & 1.04/1.03 \\
HDE~319703~A    & 17:19:46.16 & -36:05:52.4   & O7\,V((f))      + O9.5\,V     & MA16 & ...  & ...  & ...  & ...   \\
HD~159176       & 17:34:42.49 & -32:34:54.0   & O7\,V((f))      + O7\,V((f))  & S14  & 0.34/0.42 & 0.39/0.38 & 0.36/0.32 & 0.94/0.75  \\
HD~161853       & 17:49:16.56 & -31:15:18.1   & O8\,V(n)z       + B           & S14  & 0.66 & 0.52 & 0.80 & 1.20  \\
Herschel~36     & 18:03:40.33 & -24:22:42.7   & O7:\,V          + sec         & S14  & ...  & ...  & ...  & ...   \\
HD~165052       & 18:05:10.55 & -24:23:54.8   & O6\,Vz          + O8\,Vz      & MA16 & ...  & ...  & ...  & ...   \\
HD~165921       & 18:09:17.70 & -23:59:18.3   & O7\,V(n)z       + B0:\,V:     & S14  & 0.42 & 0.69 & 0.72 & 1.04  \\
ALS~4923        & 18:19:28.43 & -15:18:46.2   & O8.5\,V         + O8.5\,V     & MA16 & 0.36/0.33 & 0.32/0.32 & 0.39/0.28 & 1.09/0.84  \\
HD~175514       & 18:55:23.12 & +09:20:48.1   & O7\,V(n)((f))z  + B           & MA16 & ...  & ...  & ...  & ...   \\
HD~194649~AB    & 20:25:22.12 & +40:13:01.1   & O6.5\,V((f))    + sec         & MA16 & ...  & ...  & ...  & ...   \\
Cyg~OB2-73      & 20:34:21.93 & +41:17:01.6   & O8\,Vz          + O8\,Vz      & MA16 & 0.39/0.28 & 0.27/0.27  & 0.49/0.32 & 1.26/1.14  \\
HD~206267~AaAb  & 21:38:57.62 & +57:29:20.6   & O6.5\,V((f))    + O9/B0\,V    & S11a & 0.52 & 0.75 & 0.53 & 0.71  \\
BD~+55~2722~C   & 22:18:59.88 & +56:07:18.9   & O7\,V(n)z       + B         & MA16 & 0.51 & 0.72 & 0.80 & 1.11  \\
HD~215835  & 22:46:54.11 & +58:05:03.5   & O5.5\,V((f))    + O6\,V((f))  & S14  & 0.13/0.37 & 0.50/0.40 & 0.41/0.38 & 0.81/0.96  \\
ALS~12688       & 22:55:44.94 & +56:28:36.7   & O5.5\,V(n)((fc))+   B         & MA16 & 0.35 & 0.64 & 0.50 & 0.79  \\
\label{list-sb2}
\enddata
\end{deluxetable}

\end{landscape}

\begin{landscape}
\begin{deluxetable}{ccclc}
\tablewidth{0pt}
\tabletypesize{\footnotesize}
\tablecaption{Clusters  and H~{\sc ii} regions containing O\,Vz stars}
\startdata
\hline
  \multicolumn{1}{c}{Nebular Complex/Association} &
  \multicolumn{1}{c}{Cluster/H~{\sc ii} region} &
  \multicolumn{1}{c}{O\,Vz} &
  \multicolumn{1}{l}{Stars} & 
  \multicolumn{1}{c}{O\,V non-z} \\
\hline
  NGC~3372/Car OB1 & Trumpler 14              & 4 & {\bf CPD -58 2611}, HD 93128,          & 1 \\
                   &                          &   & HD 93129 B, {\bf CPD -58 2620}          &   \\ 
                   & Trumpler 16              & 5 & {\bf HDE 303311}, CPD -59 2591,        & 7 \\
                   &                          &   & HD 93161A, V662Car, {\bf HDE 303316 A} &    \\
                   & Collinder~228            & 4 & {\bf HDE 305524}, HD 305438,           & 1 \\
                   &                          &   & {\bf HDE 305532}, {\bf CPD -59 2673}   &    \\
                   & Bochum 11                & 2 & HDE 305539, HDE 305612                 & 0 \\
                   & NGC 3324                 & 1 & HD 92206 C                             & 1 \\
                   & NGC 3293                 & 2 & HD 91824  {\bf HD 91572}               & 0  \\
                   & ...                      & 1 & {\bf 2MASS J10224096-5930305}          & 2 \\ 
%                   & Total          &     & 20 & \\
\hline 
 Cygnus X          &  Sh 2-109             & 12 & {\bf Cyg OB2-29}, Cyg OB2-25A, ALS 15134,  & 9 \\
                   &                       &    & {\bf BD +45 3216 A}, {\bf HDE 228759},&    \\
                   &                       &    & {\bf HDE 227018}, {\bf HDE 227245}, {\bf HD 199579}    &   \\
                   &                       &    & ALS 18 707, 2MASS J20315961+4114504,   & \\ 
                   &                       &    &  {\bf Schulte 73} (both components)    & \\
 \hline
Cyg OB9            & Sh 2-108              & 1 & BD +40 4179 & 4\\ 
\hline
``Heart and Soul'' Nebula & IC 1795, IC 1805, IC 1848  & 3 &{\bf BD +61 411 A}, {\bf BD +60 501}, {\bf BD +60 586 A} & 4 \\
\hline
 Ser OB1 & NGC 6611/M16   & 3 & {\bf HD 168137 AaAb},  HD 168504, ALS 15360 & 0\\
         & NGC 6618/M17   & 1 & {\bf BD -16 4826} & 1 \\
\hline
 Sct OB3        & Sh 2-50   & 1 & V479 Sct    & 1 \\
\hline
 Cepheus/Cep~OB1, Cep~OB2  & Sh 2-145     & 2 & BD +62 2078, {\bf HD 213023 A}      &0\\
 Cep~OB3, Cep~OB4          & Teutsch 127  & 2 & BD +55 2722 A, BD +55 2722 C  & 1\\
                           & NGC 7380     & 1 & {\bf ALS 12619}                    & 1\\
                           & Sh 2-155     & 1 & {\bf HD 217086}                     & 1 \\
                           & Berkeley 59  & 2 & {\bf BD +66 1675}, {\bf Tyc 4026-00424-1} & 1 \\
\hline 
NGC 6357                   & Pismis 24    & 2 & {\bf ALS 16052}, Pismis 24-15 & 4\\
\hline
Cat's Paw Nebula           & NGC 6334     & 1 &HDE 319703 BaBb & 1\\
\hline 
  Mon OB1                  & NGC 2264     & 2 & {\bf Tyc 0737-01170-1}, 15 Mon AaAb & 0\\
\hline
  Rosette Nebula/Mon~OB2   & NGC 2244     &1 & {\bf HD 46150} & 3\\
                           & Sh 2-280     &1 & HD 46573 & 0 \\
\hline
 Gem OB1                   & NGC 2175     & 1 & {\bf HD 42088} & 0\\
\hline
  Pup OB1                 & NGC 2467      &1 & {\bf HD 64568} & 2\\
\hline 
  Sgr OB1                 & NGC 6514      & 1 & {\bf HD 164492 A} & 0 \\
\hline
 Sgr OB5                  & Sh 2-15       & 1 & {\bf HD 161 853} & 0 \\
\hline
  Vul OB1,Vul OB4         & NGC 6823      & 2 & HDE 344784 A, HDE 344777 & 0 \\
                          & Sh 2-88       & 1 & HDE 338916              & 0 \\ 
\hline 
  Aur OB2,Aur OB1         & NGC 1893      & 5 & ALS 8294, BD +33 1025 A, {\bf HDE 242908}, & 0 \\ 
                          &               &   & HDE 242926, HDE 242935 A & \\
\hline
  CMa OB1                 & IC 2177,Sh 2-296 & 1 &HD 53975 & 0\\
\hline
  Cam OB1,Per OB3,Cam OB3 & Alicante 1       & 1 & {\bf ALS 7833} & 1\\
\hline
Sco OB1                   & Trumpler 24      & 1 & {\bf HD 152590} & 0 \\
\hline
Cas OB7                   & Sh 2-180         & 1 & ALS 6351 & 0 \\
\hline
Collinder 121             & Sh 2-306         & 1 & ALS 458     & 0 \\
                          & ...              & 1 & BD -15 1909 & 0\\
\hline
...                         & RCW 113          & 1 & HDE 326 775 & 1\\ 
\hline
...                       & RCW 54           & 1 & {\bf HD 96715} & 2 \\
\hline
...                    & ...              & 3 & {\bf HD 44811}, {\bf HD 97966}, HD 19265 & 12 \\
%\hline
\label{list-cumulos}
\enddata
\end{deluxetable}

\end{landscape}

\begin{landscape}
\begin{deluxetable}{cccl}
\tablewidth{0pt}
\tabletypesize{\footnotesize}
\tablecaption{Clusters containing O\,V stars but not Vz objects}
\startdata
\hline
  \multicolumn{1}{c}{Nebular Complex/Association} &
  \multicolumn{1}{c}{Cluster/H~{\sc ii} region} &
  \multicolumn{1}{c}{O\,V} &
  \multicolumn{1}{l}{Stars} \\
\hline
 ...     & Pismis 11     & 1 & CPD -49 2322   \\
 ...     & Bochum 1      & 1 & HDE 256725 A \\
 Car OB1 & Bochum 9      & 1 & HD 91837 \\
 Sgr OB1 & Collinder 367 & 1 & HD 165921  \\
 ...     & Havlen-Moffat 1 & 2 & ALS 18770, ALS 18768 \\
 ...     & IC 1590     & 2 & HD 5005 A, HD 5005 C \\
Cru OB1  & IC 2944     & 3 & HD 101191, HD 101223, HD 101413 AB \\
Sgr OB1  & IC 4701     & 1  & HD 167633 \\
 ...     & NGC 6383    & 1  & HD 159176 \\
Ser OB2  & NGC 6604    & 2  & ALS 4880, HD 168461 \\
 ...     & RCW 114     & 1  & HD 155913 \\
Ser OB1  & RCW 161     & 1  & ALS 4923 \\
...      & RCW 167     & 1  & BD -10 4682 \\
...      & Sh 2-10     & 1  & Tyc 7370-00460-1 \\
Cas OB7, Cas OB1 & Sh 2-185      & 1 & BD +60 134 \\
Cas OB 6         & ...           & 1 & BD +62 424 \\
Aur OB1          & Sh 2-227  & 1 & ALS 8272 \\
Aur OB2, Aur OB1 & Sh 2-230  & 1 & HD 35619 \\
...              & Sh 2-289  & 1 & LS 85 \\
Ser OB1          & Sh 1-48   & 1 & BD -14 5014\\
Cep OB2                & Trumpler 37  & 1 & HD 206267 AaAb\\
Vel OB1, Vel OB2       & ...& 1 & V467 Vel\\
Cas OB5                & ...& 1 & BD +60 2635\\
...                    & ...& 1 & BD +55 2840\\
Cep OB2              & ...& 1 & ALS 12050\\
Vul OB1, Vul OB4     & ...& 1 & HDE 344758\\
Cen OB1              & ...& 1 & HD 110360 \\
%\hline
\label{list-non-z}
\enddata
\end{deluxetable}

\end{landscape}
%********************************************************************

\acknowledgments

We thank the anonymous referee for useful suggestions and comments on the
manuscript. 
We also thank the staff of Las Campanas Observatory for their support
during the observing runs.
J.I.A and R.H.B acknowledge financial support from the Chilean Government
grants, through the FONDECYT Iniciaci\'on 11121550 and FONDECYT Regular 1140076
projects, respectively.
S.S.-D. acknowledges financial support from the Spanish Ministry of Economy
and Competitiveness (MINECO) under grants AYA2010-21697-C05-04, and Severo
Ochoa SEV-2011-0187, and by the Canary Islands Government under grant
PID2010119. 
J.M.A., A.S., and E.J.A. acknowledge support from [a] the Spanish Government
Ministerio de Econom{\'\i}a y Competitividad (MINECO) through grants
AYA2010-15\,081, AYA2010-17\,631, and AYA2013-40\,611-P and [b] the
Consejer{\'\i}a de Educaci{\'o}n of the Junta de Andaluc{\'\i}a through grant
P08-TIC-4075. C.S.-S. acknowledges support from the Joint Committee
ESO-Government of Chile. 
I.N. and A.M. acknowledge support from the Spanish Ministerio de
Ciencia e Innovaci\'on (MICINN) under
grant AYA2012-39364-C02-02, and by the Generalitat Valenciana (ACOMP/2014/129).
A.M. acknowledges support from the Generalitat Valenciana through the grant
BEST/2015/242. 
The Space Telescope Science Institute is operated by the Association of
Universities for Research in Astronomy, Inc., under NASA contract NAS5-
26555. 
This research has made use of Aladin (Bonnarel et al. 2000), and
the SIMBAD database, operated at CDS, Strasbourg, France.

\end{document}